\documentclass{article}
\usepackage[utf8]{inputenc}
\usepackage{amsmath}
\usepackage{graphicx}
\usepackage[a4paper, total={6in, 8in}]{geometry}
\usepackage{listings}
\usepackage{float}
\usepackage{placeins}
\usepackage[nottoc,numbib]{tocbibind}
\usepackage{xcolor}
\usepackage{hyperref}
\hypersetup{colorlinks,linkcolor={blue},citecolor={blue},urlcolor={blue}}
\usepackage[]{lineno}
\usepackage{bm}
\usepackage{url}
\usepackage[affil-it]{authblk}
\usepackage{subfig}
\providecommand{\keywords}[1]{Keywords: #1}

\title{Numerical study on wave attenuation
via 2D fully kinetic electromagnetic
particle-in-cell simulations
}

\author[1]{Fei Du}
\author[1]{Yize Yan}
\author[1]{Jingfeng Tang}
\author[1]{Daren Yu}
\author[1,*]{Yinjian Zhao}

\affil[1]{School of Energy Science and Engineering,
Harbin Institute of Technology,
Harbin 150001, People’s Republic of China}
\affil[*]{Corresponding author: Yinjian Zhao, zhaoyinjian@hit.edu.cn}

\date{\today}

\begin{document}

\maketitle

\begin{abstract}
The propagation and absorption of electromagnetic waves in plasma is one of the fundamental issues in plasma physics. The electromagnetic particle-in-cell method with the finite-difference time-domain solver plus Monte Carlo collision model would be the most accurate method to simulate the wave-plasma interaction. However, the numerical effects of this method have not been carefully investigated especially in two dimensions. In this paper, the 2D PIC method is used to study the electromagnetic wave attenuation by fluorescent lamp plasma tubes. 
The study finds that the number of macro-particles and the incident electromagnetic wave amplitude have minor effects on the wave attenuation within a certain appropriate parameter range. Furthermore, the effects of electromagnetic wave frequency, the plasma distribution structures, and collision types on wave attenuation are investigated. Particularly,
it is found that the staggered way of arranging the plasma distribution structures  can achieve better wave attenuation than the parallel way, which agrees with our recent experimental observation.


\end{abstract}

\section{Introduction}
Plasma plays a crucial role in the interaction with electromagnetic waves in various scientific and technological fields. As electromagnetic waves propagate through plasma, they experience refraction, reflection, and attenuation. These characteristics are useful in the development of materials for absorbing electromagnetic waves, antenna design, and wireless communication modulation techniques\cite{SJES6AFFAE5AEEC7647980C0AA1D7ED0010B,XJAZB4C513E52A39B80C506EC24707BC1448}. The study of plasma and electromagnetic waves has revealed its physical properties, offering broad application prospects in fields such as telecommunications.

In the $1950$s and $1960$s, driven by military needs, both domestically and internationally, a substantial amount of theoretical research was conducted on the interaction between plasma and electromagnetic waves. Over the decades, significant progress has been made in theoretical research. Ginzburg, Vidmar, and others established a series of theoretical models for the interaction between electromagnetic waves and plasma \cite{57528,1970The}, which have been widely applied by subsequent scholars. The development of computers has also led to great advancements in the numerical simulation of the interaction between plasma and electromagnetic waves.
In $2003$, Shaobin Liu and his colleagues employed the WKB method to study the influence of electron density, collision frequency, and electromagnetic wave frequency on the attenuation of electromagnetic waves in plasmas with linear and exponential distributions\cite{liu2006wkb,shaobin2008wentzel}. The WKB method, which treats stratified plasmas through geometric optics approximation and requires each layer to be uniform with minimal parameter variations, is applicable to solving electromagnetic wave propagation in one-dimensional plasmas limited in simple structures.

When the simulation object is two-dimensional or even three-dimensional, the main applicable methods are the finite-difference time-domain (FDTD) method. In $2009$, Chaudhury and his team used the Finite-Difference Time-Domain (FDTD) method to study the Radar Cross Section (RCS) of a metal plate covered with a plasma with three-dimensional inhomogeneity. The frequency of the incident electromagnetic wave was in the range of $3$-$9$ GHz. They found that the attenuation of the electromagnetic wave by the plasma was related to the angle of incidence\cite{2005Three,2006Comparison,2009Study}.
However, the FDTD method uses complex permittivity to approximate all plasma behaviors, which cannot resolve detailed particle motion and trajectories. In contrast, the particle-in-cell (PIC) method combines the FDTD method to solve the electromagnetic fields and tracks the motion of plasma particles by solving the equations of motion. In the PIC method, the plasma is naturally represented in the form of macro-particles and is connected to the Maxwell's equations through the charge and current density deposited by the particles at grid points. Moreover, collisions between charged particles can be handled using the Monte Carlo collision (MCC) model. Therefore, if computational resources permit, using PIC combined with the MCC model and the FDTD solver would be the most accurate method for studying the interaction between electromagnetic waves and plasmas.

\begin{figure}[htb]
	\centering
	\includegraphics[width=0.48\linewidth]{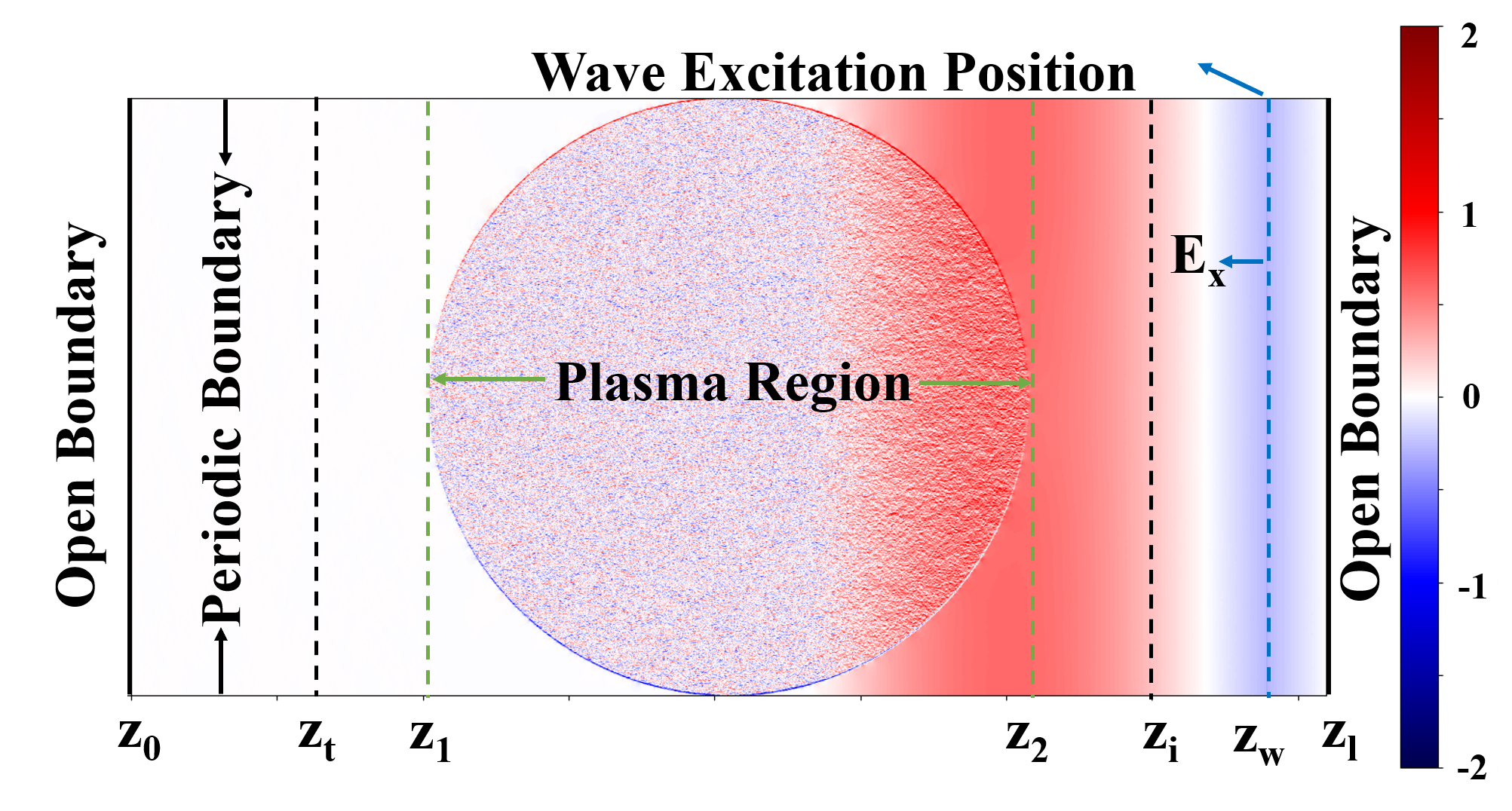}
        \caption{Illustration of the simulation setup.}
	\label{setup}
\end{figure}

The early application of the PIC method can be traced back to 1955 when researchers at the Los Alamos National Laboratory in the United States developed this method to simulate the behavior of charged particles in plasmas \cite{fermi1955studies,harlow1988pic}. Numerous studies have employed the PIC method to investigate wave-plasma interactions in various applications, but relatively few studies have focused on using the PIC method to research the attenuation process of electromagnetic waves. In $2014$, the pioneering work by Yanxia Xu et al\cite{xu2013pic}. applied the PIC-MCC method to study electromagnetic wave attenuation. The study found that the power attenuation of electromagnetic waves is strongly influenced by plasma density, neutral gas species, neutral gas density, and electromagnetic wave frequency. In $2022$, Dongning Gao et al\cite{gao2022attenuation}.  used the PIC method to study the attenuation rates of electromagnetic waves with different wavelengths in plasmas. It was evident that the attenuation rate increases with the wavelength of the electromagnetic wave; when the wavelength is greater than $0.07$ meters, the attenuation rate slowly increases and eventually shows little to no increase.

From existing theoretical and experimental research, we know that to achieve the optimal absorption effect of electromagnetic waves, it is necessary to allow electromagnetic waves to enter the plasma as much as possible and maximize the propagation distance of electromagnetic waves within the plasma to achieve minimal reflection of electromagnetic signals. In order to enhance the high absorption of plasma for electromagnetic waves, as well as to meet the demands of increasingly advanced radar detection technology and communication control technology, scholars have studied the transmission performance of electromagnetic waves in plasmas with different spatial arrangements\cite{57528,xuesong2022investigation}. 
\begin{figure}[htbp]
	\centering
	\includegraphics[width=0.43\linewidth]{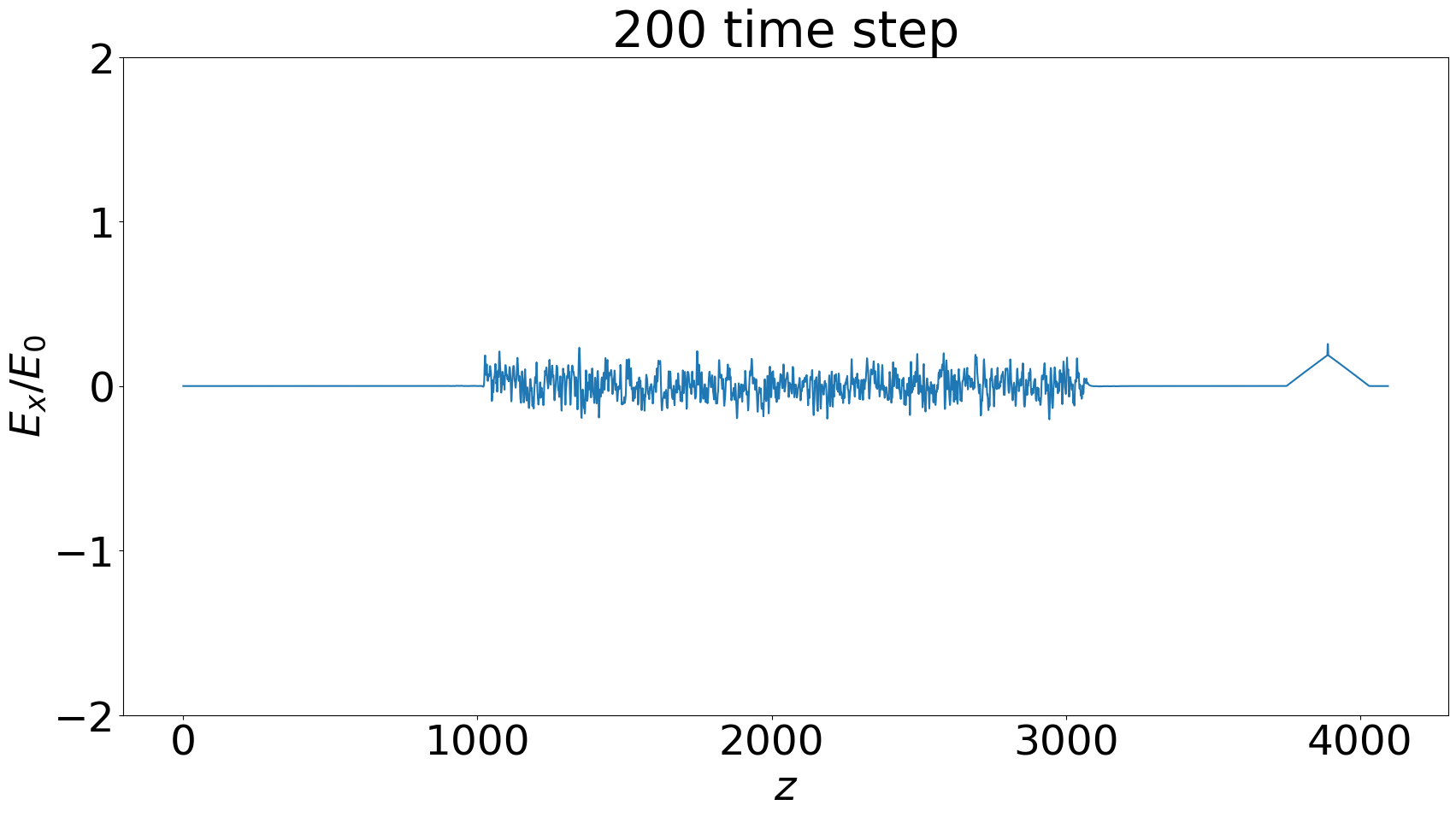} 
	\includegraphics[width=0.43\linewidth]{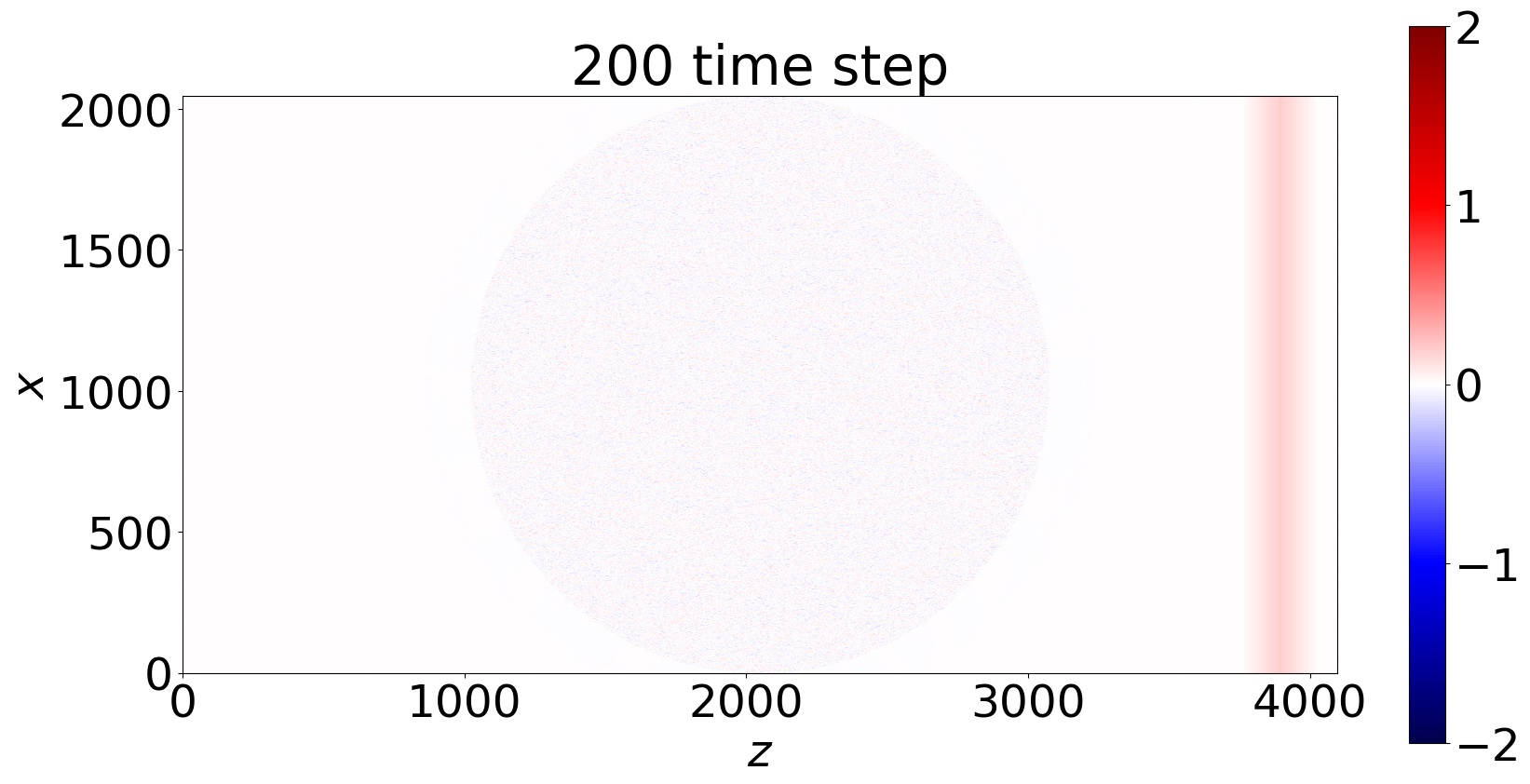} \\
	\includegraphics[width=0.43\linewidth]{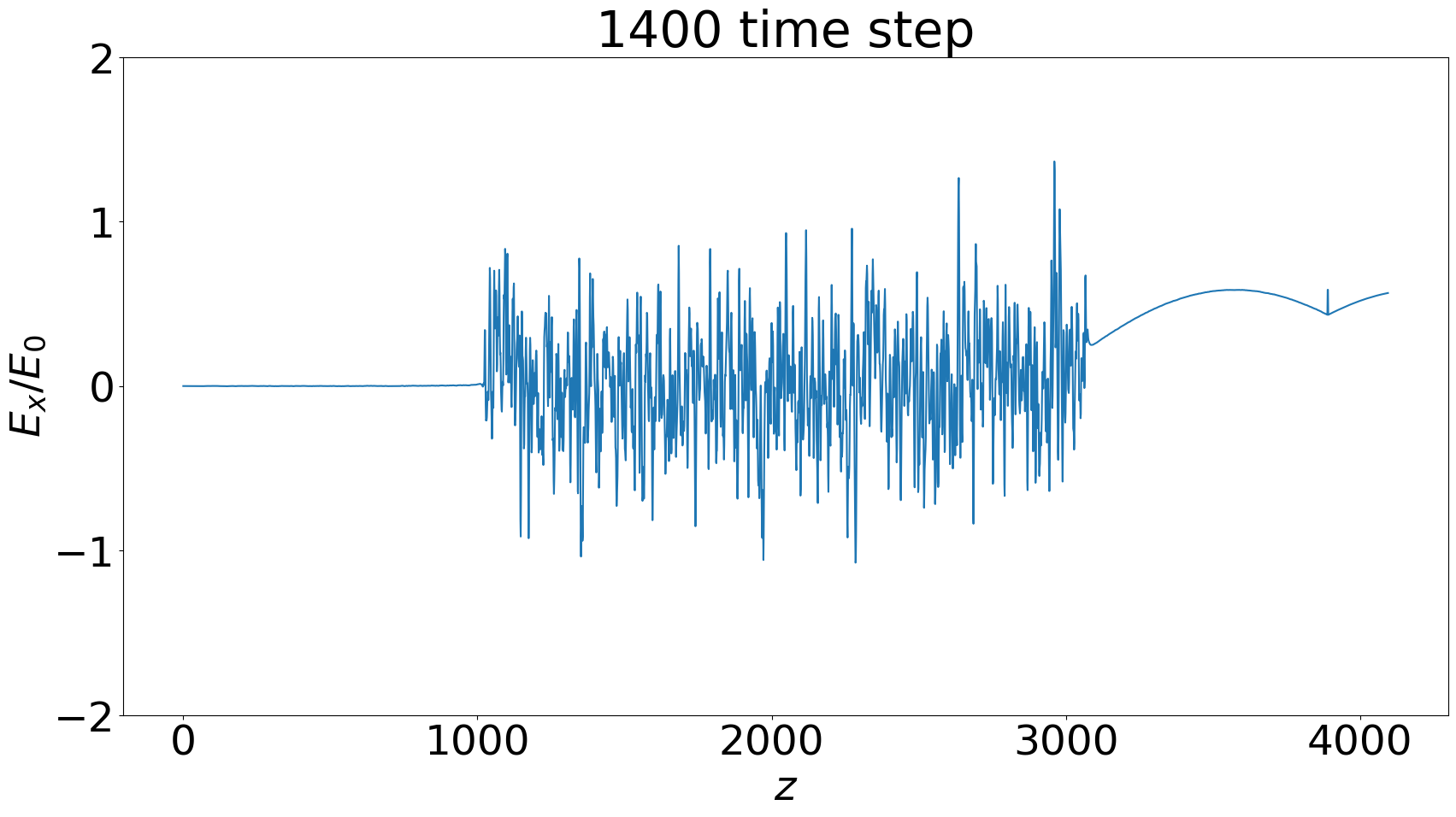} 
	\includegraphics[width=0.43\linewidth]{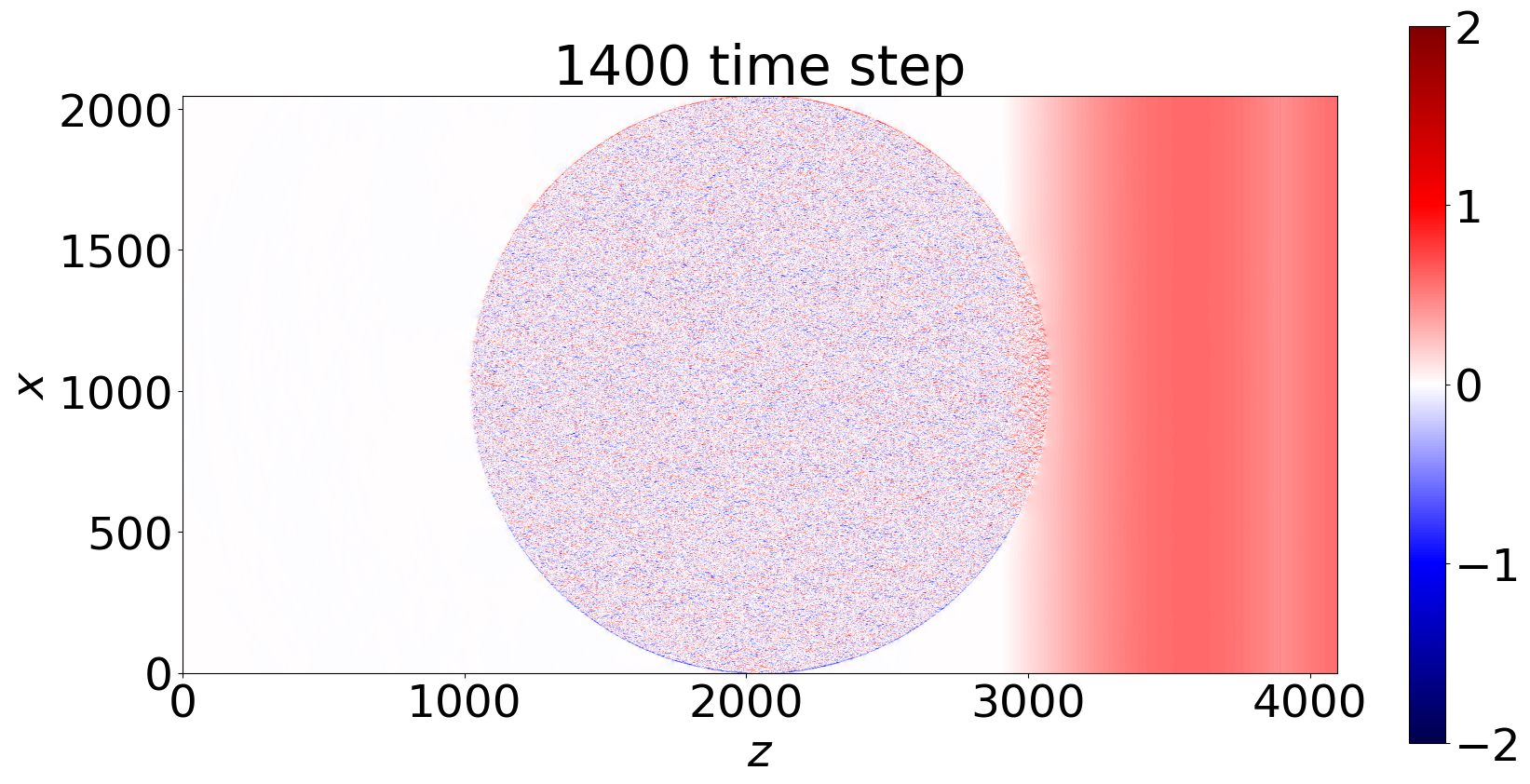} \\
	\includegraphics[width=0.43\linewidth]{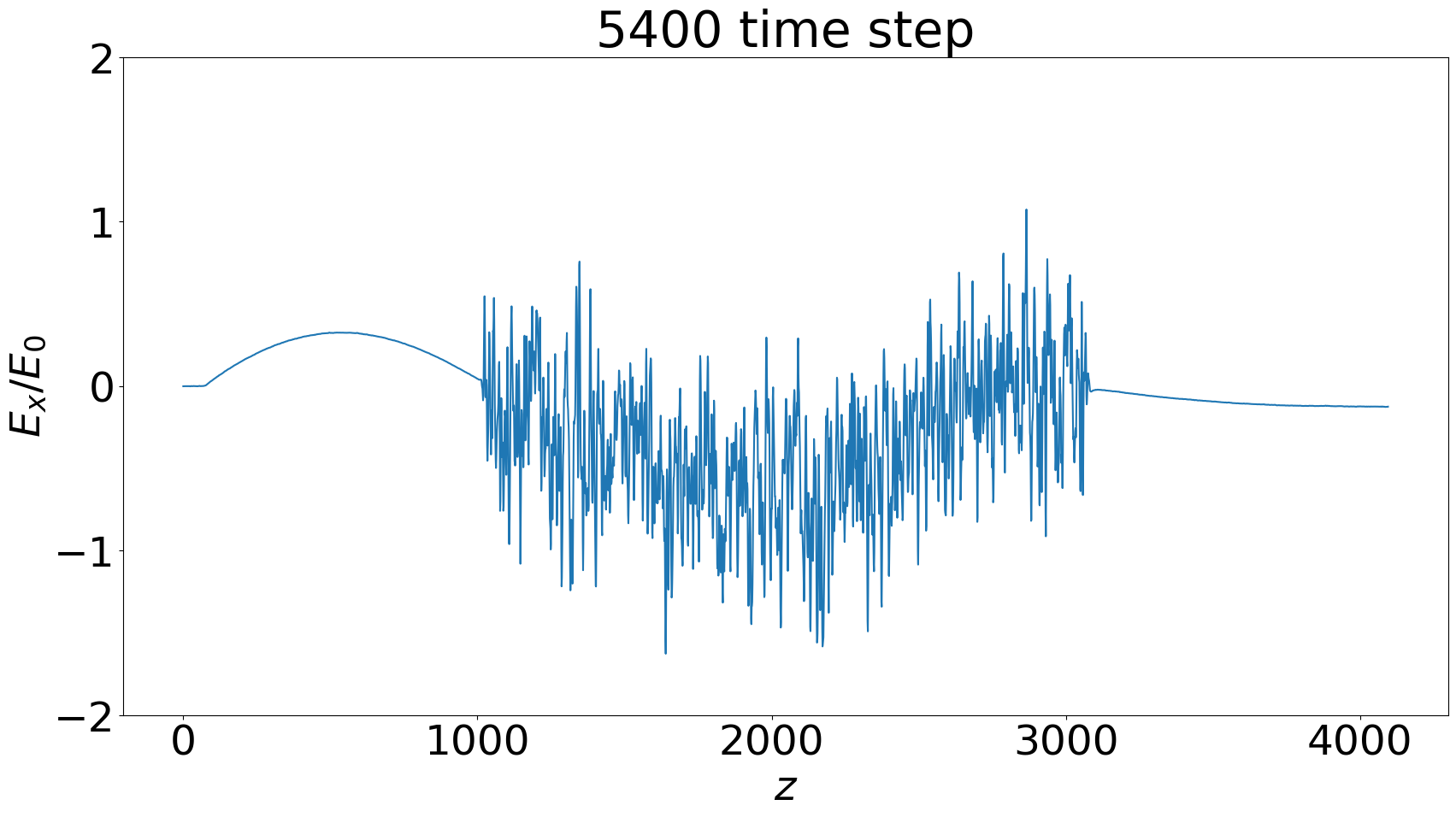} 
	\includegraphics[width=0.43\linewidth]{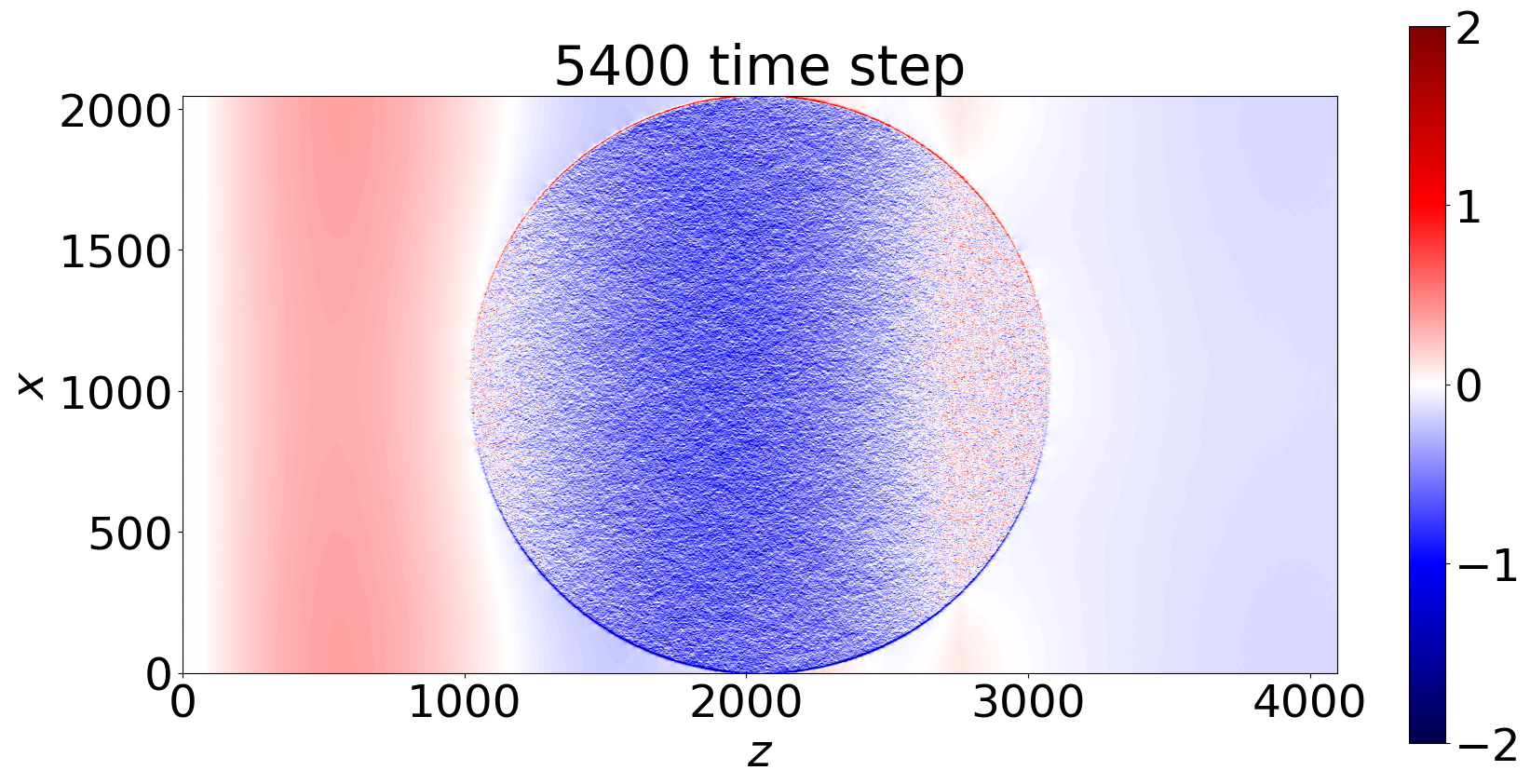} \\
	\includegraphics[width=0.43\linewidth]{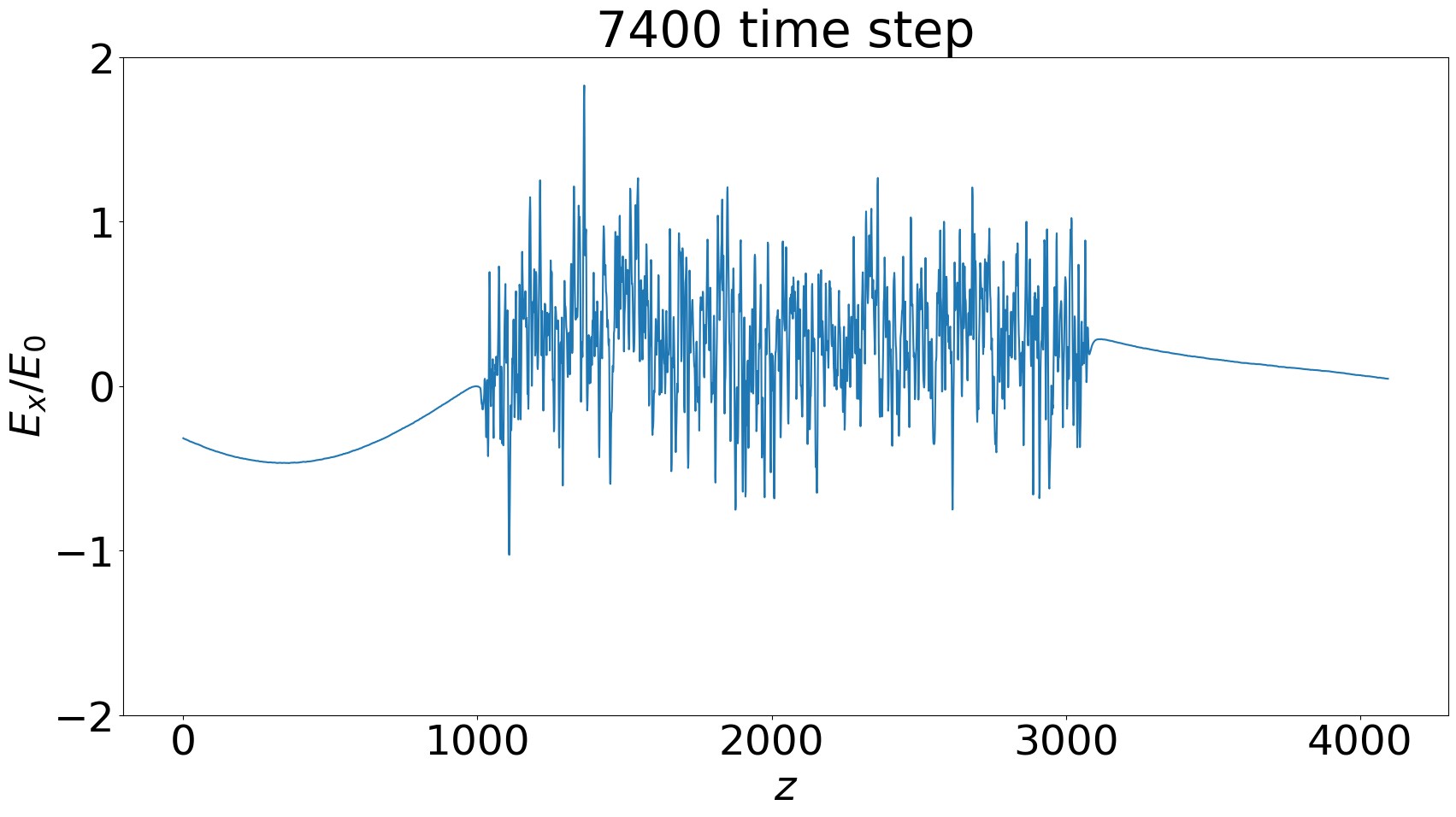} 
	\includegraphics[width=0.43\linewidth]{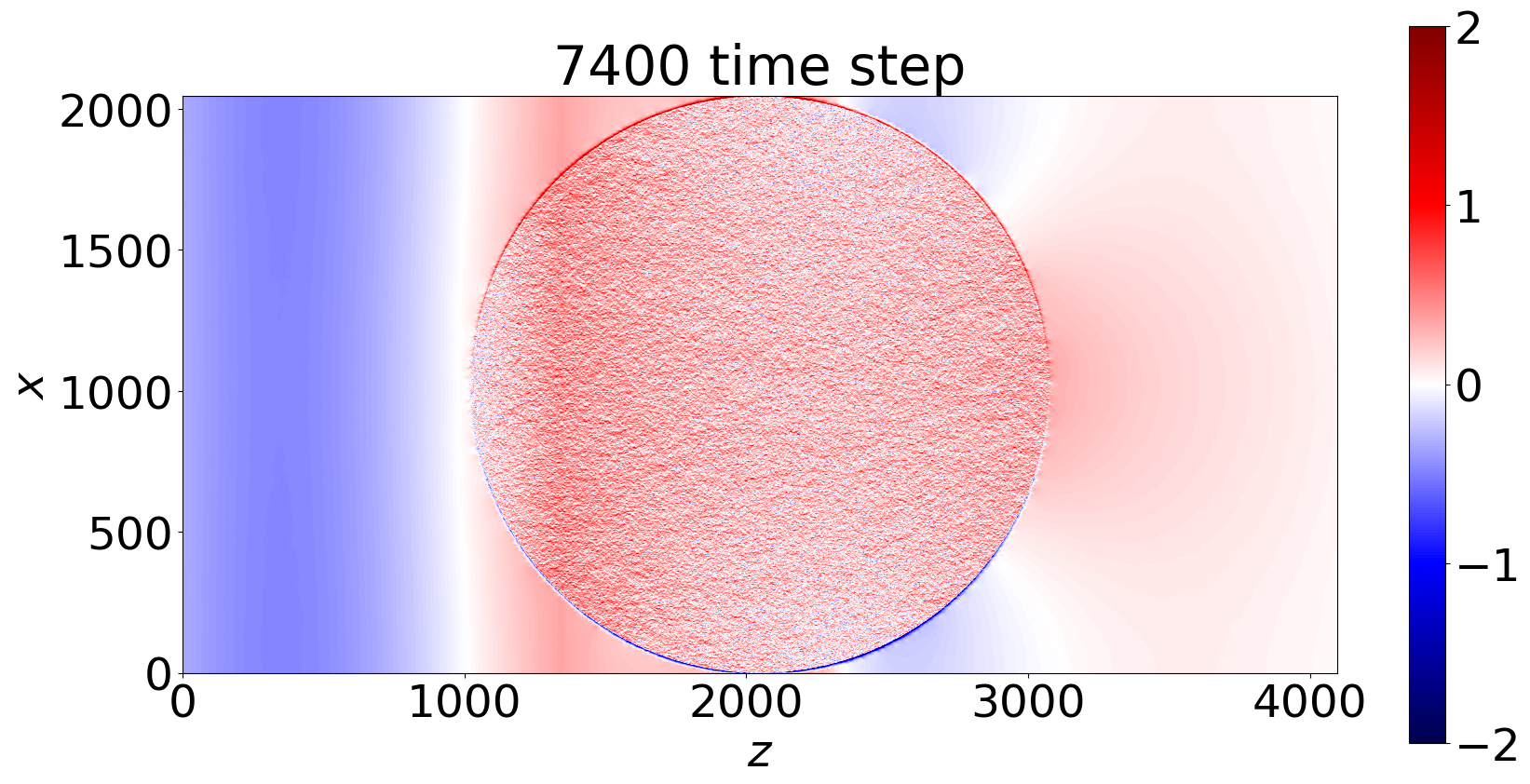}
	\caption{The propagation process of electromagnetic waves.}
	\label{Ex}
\end{figure}

The fluorescent lamp plasma tube, as a commonly used plasma generation device, is readily accessible and offers relatively simple control over plasma parameters. After years of development, the fluorescent lamp plasma tube has received extensive attention in the field of wave attenuation research and has achieved significant accomplishments\cite{harlow1988pic,anderson2007plasma,danilov1997electromagnetic}. For example, Danilov et al\cite{1997Electromagnetic}. studied the reflection, transmission, and absorption coefficients of periodically arranged plasmas for electromagnetic waves in the $4$-$12$ GHz range. Howlader et al\cite{2005Time}. investigated the attenuation of $10$ GHz electromagnetic waves by fluorescent tube plasma and used the microwave interferometry method to determine that the electron number density of the fluorescent lamp plasma ranges from $3.5 \times 10^{10} \text{ to } 2.6 \times 10^{11} \text{ cm}^{-3}$, with collision frequencies between $7$ and $40$ GHz. The Cappelli team at Stanford University arranged gas discharge tubes in a periodic three-dimensional ``stake'' structure to form a tunable plasma photonic crystal. Their research concluded that the transmission effect with all discharge tubes open is different from the superposition of the transmission effects in the vertical and horizontal directions, demonstrating that the influence of different spatial structural positions on electromagnetic wave transmission is not the same and is mutually coupled\cite{SIPD42CCED5A681F923759E40183B30D73C4}. Although some progress has been made in the study of wave attenuation in plasma discharge tube arrays, to our knowledge, there are few papers that simulate and study the attenuation of electromagnetic waves in plasma discharge tubes with different structural distributions using 2D PIC methods. Therefore, considering the range of plasma parameters, reliability, controllability of the discharge state, cost-effectiveness, and the convenience of constructing equipment arrays, and referring to the existing literature on such devices, this paper selects the plasma discharge tube as the two-dimensional simulation geometry.

In the aforementioned work applying the Particle-in-Cell (PIC) method to study wave-plasma interactions, the PIC method is directly used as a tool for investigating physical problems. However, the detailed numerical impact of PIC parameters on wave-plasma interactions has not been reported or deeply explored. Simultaneously, studies on plasma arrays with simple configurations have demonstrated that the arrangement structure of the plasma has a significant effect on the attenuation efficiency, capable of enhancing absorption. A previous work of us in $2024$, Yan et al. \cite{yan2024numerical}, studied the effects of numerical and physical parameters, such as the electromagnetic wave frequency, amplitude, plasma temperature, thickness, and collision types on wave attenuation through one dimensional fully dynamic electromagnetic PIC simulations. However, only the absorption effects of one-dimensional distributed plasma have been investigated, lacking a two-dimensional PIC study on the absorption of electromagnetic waves by different plasma distributions.

As far as we know, there are currently few research papers on the study of electromagnetic wave attenuation in plasma using two-dimensional ($2$D) Particle-in-Cell (PIC) simulations. Therefore, in this paper, we employ a two-dimensional electromagnetic PIC with Monte Carlo Collisions (MCC) to study the wave-plasma interaction problem. In Sec.\ref{sec:2}, the numerical methods are described, including the applied PIC code and considered simulation setup, the characteristic simulation results and diagnostics , and the numerical effects of PIC parameters are examined. In Sec.\ref{sec:3}, the influence of numerical and physical parameters is studied, including the number of macro-particles, EM wave frequency, amplitude, collisions, and plasma distribution structures. At last in Sec.\ref{sec:4}, conclusions are drawn and future research plans are described.

\section{Numerical method}
\label{sec:2}

\subsection{The applied PIC code and algorithms}

In this study, the simulation tool we employed is the WarpX open-source PIC code \cite{warpx}, a highly parallelized and optimized code capable of scaling on some of the world's largest supercomputers, and in $2022$, it was awarded the ACM Gordon Bell Prize. Although WarpX was initially developed for plasma-based accelerator applications \cite{ZhaoPOP2020,ZhaoPOP2022}, it has now been extended to a variety of plasma simulation fields, including strong-field quantum electrodynamics \cite{Fedeli_2022}, relativistic magnetic reconnection \cite{Klion_2023}, plasma propulsion \cite{Wang_2023,xie2024effect}, and more. WarpX features a range of capabilities, supporting $1-3$D Cartesian and $2$D cylindrical (RZ) coordinate systems, and includes electromagnetic field solvers such as FDTD (Finite-Difference Time-Domain), CKC (Current-Conserving, Charge-Conserving algorithm), PSATD (Pseudo-Spectral Analytical Time-Domain), and MLMG (Multi-Level Multi-Grid) electrostatic field solver. Additionally, it integrates modules for Coulomb collisions and ionization Monte Carlo collisions, supports adaptive mesh refinement, Perfectly Matched Layers (PML) and other absorbing boundary conditions, Lorentz boosted frame techniques, embedded boundary handling, and efficient CPU/GPU parallel domain decomposition techniques. 
\begin{figure}[htbp]
	\centering
	\subfloat[The transmitted power during the first half-wave period.]{
		\includegraphics[width=0.48\linewidth]{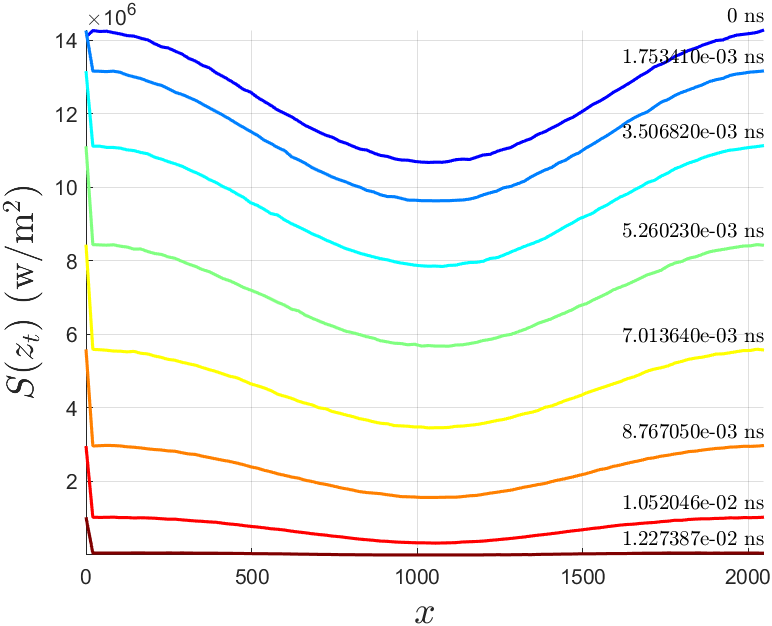}
		\label{zt_1}
	} \hfill
	\subfloat[The transmitted power during the second half-wave period.]{
		\includegraphics[width=0.48\linewidth]{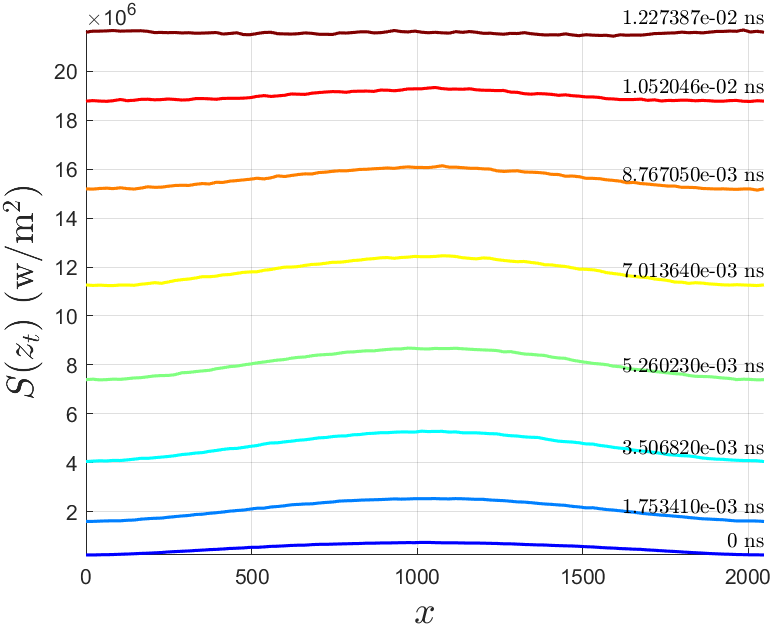}
		\label{zt_2}
	}
	\caption{The Poynting Vector at $z_t=800$.}
	\label{sum:zt}
\end{figure}
To study the process of interaction between incident electromagnetic waves and plasma, as well as the interaction of these waves with electrons and ions in the plasma, we selected the $2$D geometric configuration of WarpX, employed the classical FDTD electromagnetic field solver, and Perfectly Matched Layers absorbing boundary and the Monte Carlo ionization collision algorithm. 
These settings enable us to self-consistently resolve the interaction between the incident electromagnetic waves and those induced by the plasma, as well as their interactions with electrons and ions of the plasma. 

\subsection{Simulation setup}

The simulation setup is shown in Fig.\ref{setup}. The simulation domain is a two-dimensional space ranging from $z_0$ to $z_l$,
The $z$ boundaries are equipped with PML open boundary conditions for electromagnetic waves, allowing waves to be absorbed when they reach these boundaries; the $x$ boundaries have periodic boundary conditions, where electromagnetic waves entering these boundaries will be periodically mapped to the opposite side of the grid. The colorbar on the right side of the graph represents the normalized $E_x$, with different colors indicating different data values. The incident electromagnetic wave is excited at $z_w$ and propagates in the $\pm z$ directions. Waves propagating along $z$ will be absorbed by the boundaries, while waves propagating along $-z$ will enter the plasma. At the position $z_w$, we apply a periodic sine wave excitation, and the propagation equation for the electric field component $E_x$ can be expressed as, 
\[
E_x(t) = E_0 \sin{ (2 \pi f_0 N \Delta t)},
\]
where $N$ is the number of computational steps, $\Delta t$ is the timestep, $f_0$ is the frequency of the electromagnetic wave, and $E_0$ is the amplitude of the electromagnetic wave. According to electromagnetic wave theory, in the simulation, the electromagnetic wave is excited in two field components, namely the electric field component $E_x$ and the magnetic field component $B_y$. In the simulation, only the value of $E_x$ is set, and the corresponding $B_y$ is self-consistently resolved by the FDTD algorithm with an amplitude of $B_0 = E_0/c$, where $c$ is the speed of light.
\begin{figure}[htbp]
	\centering
 	\subfloat[Time evolution of the Poynting vectors at $z_i$]{
		\includegraphics[width=0.48\linewidth]{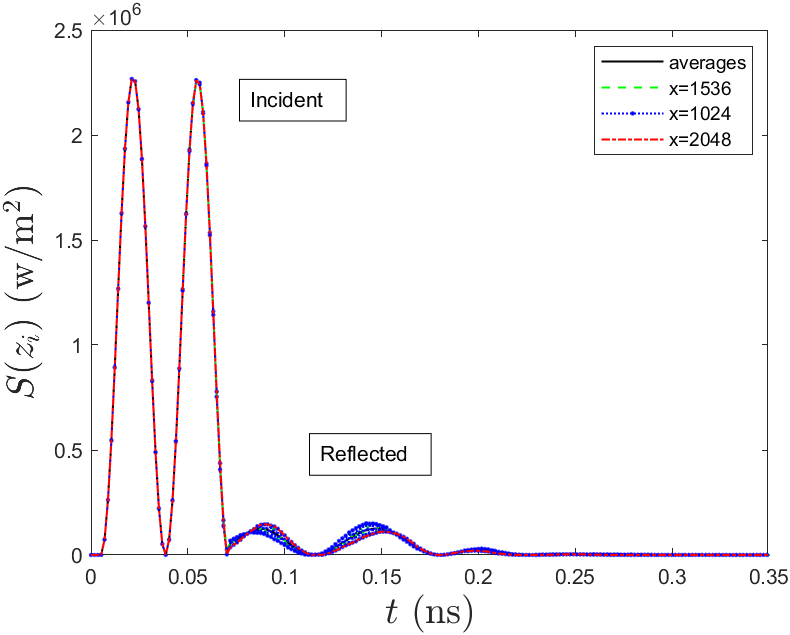}
		\label{x=zt}
	}\hfill
	\subfloat[Time evolution of the Poynting vectors at $z_t$]{
		\includegraphics[width=0.48\linewidth]{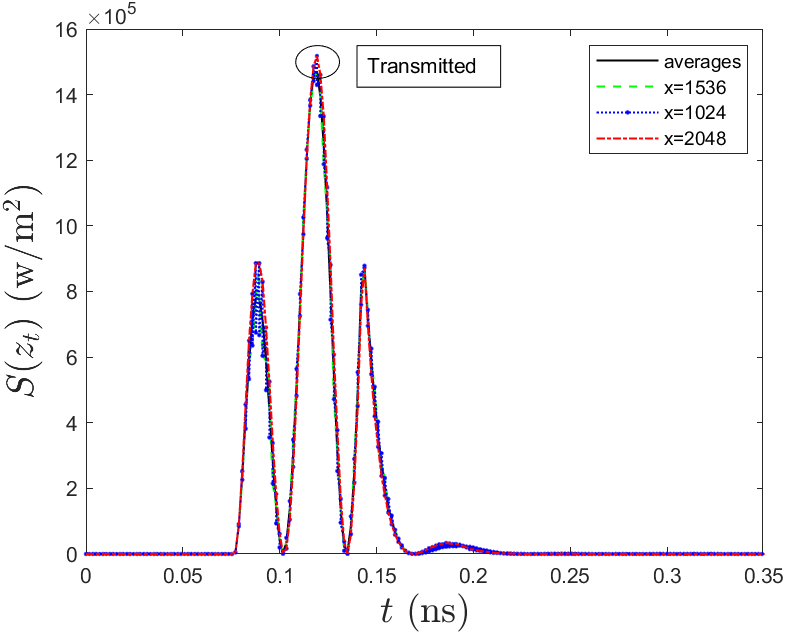}
		\label{x=zi}
	} 

	\caption{Time evolution of the Poynting vectors at Probes at  $x=1536$, $x=1024$, $x=2048$, and the Average Value.}
	\label{x=}
\end{figure}

In the plasma region between  $z_1$ and $z_2$, the particle number density is $n_0=n_e=n_i$, with particles distributed uniformly and randomly. The velocity distributions of electrons and ions follow the Maxwell-Boltzmann distribution, with temperatures of $T_e$ and $T_i$, respectively. The subscripts `$e$' and `$i$' denote electrons and ions, while `$a$' represents neutral atoms. The number density of neutral atoms is $n_a$, and their temperature is $T_a$. These parameters are used in the Monte Carlo Collision (MCC) algorithm to handle the collision processes between particles. In this study, no additional boundaries are required to be set for the plasma region, as we found that the ion motion has a negligible effect on the simulation results when the electromagnetic wave passes through the plasma, while the electrons are effectively confined within the plasma region. The ion species considered is argon, and the MCC algorithm includes the processes electron-neutral scattering, excitation and ionization for argon, which were conformed to be the most important collision types in our previous 1D study\cite{yan2024numerical}.

\subsection{Characteristic simulation results and diagnoses}

During the simulation process, we first selected a characteristic simulation with the following parameters: the grid number along the horizontal axis (z-axis) was set to $4096$, and the grid number along the vertical axis (x-axis) was set to $2048$. The plasma density was $n_0$ = $10^{18}$ m$^{-3}$, the argon density was $n_a$ = $2.4 \times 10^{23}$ m$^{-3}$. The temperatures of electrons and ions were $T_e$ = $1$eV and $T_i$ = $0.1$ eV, respectively, and the temperature of argon was $T_a$ = $0.1$ eV. The frequency of the incident electromagnetic wave was $f_0$ = $15$ GHz, which is greater than the plasma oscillation frequency $\omega_{pe}/2\pi \approx 9$ GHz. The resulting Debye length was $\lambda_D \approx 7.4$ $\mu$m, and the cell size was set to the Debye length, $\varDelta x = \varDelta z = \lambda _D$. 
In this simulation,the computational domain was $15.1552\,\text{ mm} \times 30.3104\,\text{mm}$ in size, and the value of the time step ($\Delta t$) was adaptively given by the WarpX program to meet the CFL limit, which is related to the cell size and the speed of light, and under the current computational conditions, it is set to $\Delta t$ = $1.753409952\times 10^{-14}s$.
The plasma was located between $z_1 = 7.6\text{ mm}$ and $z_2 = 22.8 \text{ mm}$, within a spherical region of radius $r_0 = 7.6 \text{ mm}$. The wave excitation position was at $z_w = 28.9 \text{ mm}$. The origin was set at $z_0 = 0 \text{ mm}$, and the domain length was $z_l$ - $z_0$ = $30.3$ mm. The probe positions were $z_i$ = $27.4$ mm and $z_t = 0.1 \text{ mm}$, which are used to record the incident and transmitted wave information, respectively. 
\begin{figure}[htbp]
	\centering
	\includegraphics[width=0.48\linewidth]{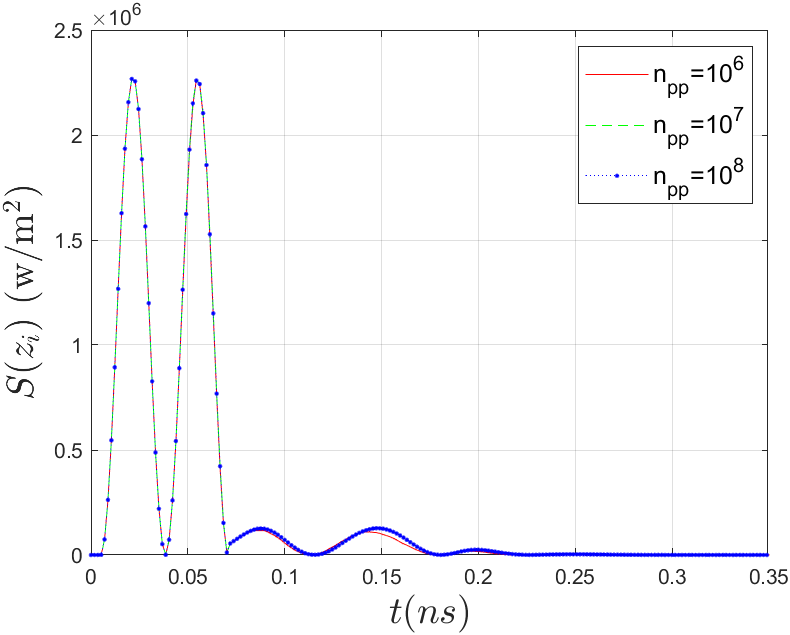} \hfill
	\includegraphics[width=0.48\linewidth]{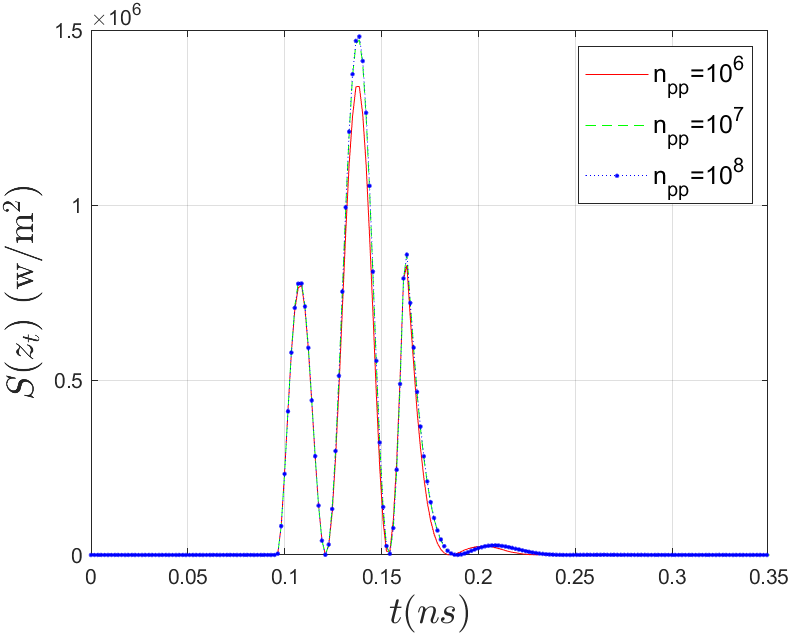}
	\caption{Time evolution of the Poynting vectors at $x_i$ and $x_t$ of the case with $n_{pp}=10^6$,$n_{pp}=10^7$ and $n_{pp}=10^8$.}
	\label{sum:npp}
\end{figure}

The simulation was run for $20000$ time steps. Most parameters were chosen based on reference\cite{yan2024numerical,Xu_2014}. $E_0 = 5 \times 10^{4}$ V/m was selected such that $E_0/n_a \approx 208$ \text{Td}, which is within a reasonable range\cite{PANCHESHNYI2012148}.
The simulation was conducted on a personal desktop computer equipped with an Intel\textsuperscript{\textregistered} Xeon\textsuperscript{\textregistered} Platinum 8269CY CPU @ $2.50$GHz. 
This computer possesses $26$ physical cores and $52$ logical processors. Throughout the simulation, $16$ MPI (Message Passing Interface) processes were employed to parallelize the simulation, aiming to enhance the efficiency and speed of the simulation.
\begin{figure}[htb]
	\centering
	\includegraphics[width=0.6\linewidth]{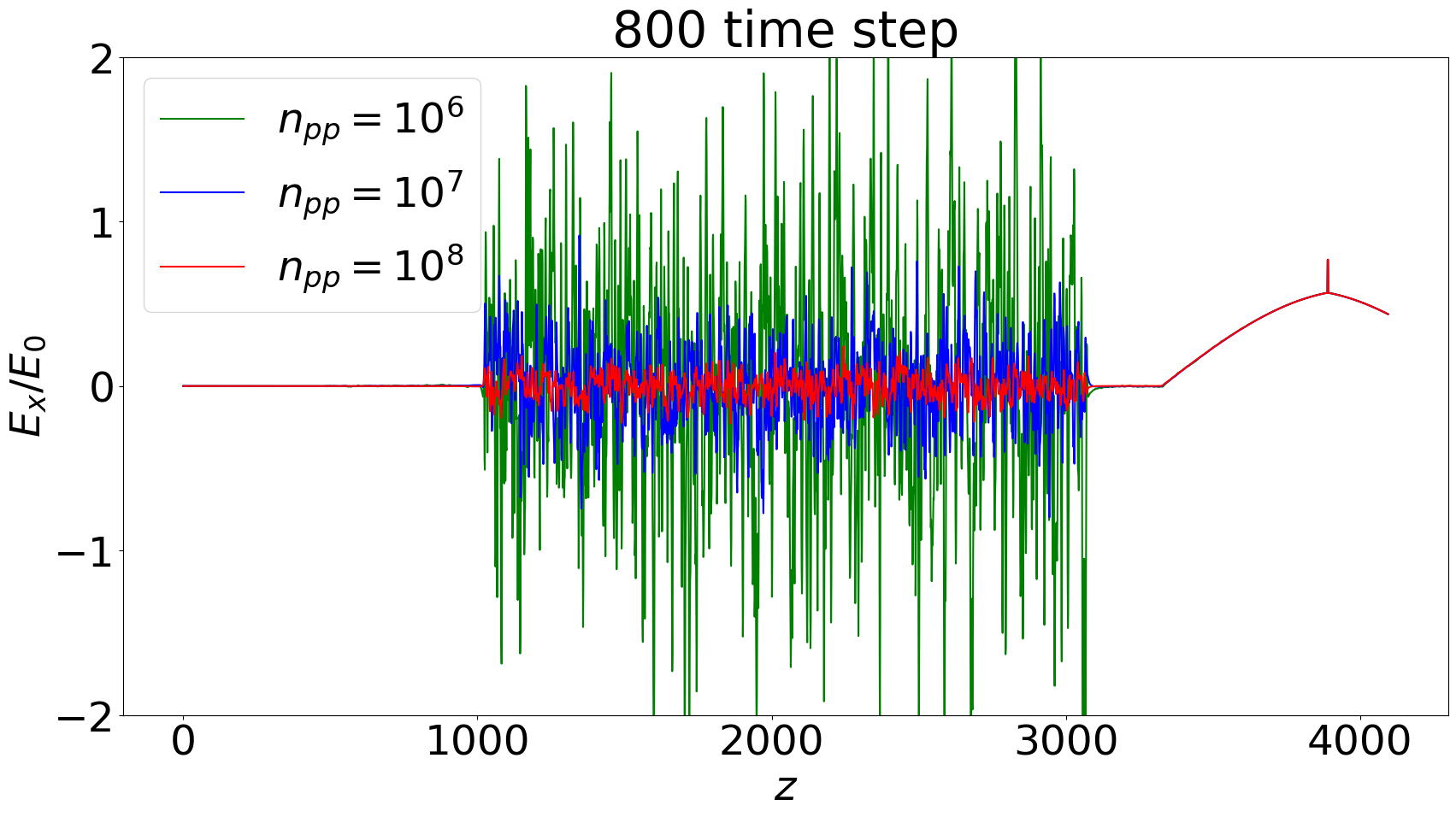}
	\caption{$1$D Electromagnetic Wave Propagation Process with Different Numbers of Particles at $800$ Steps.}
	\label{fig:npp1D}
\end{figure}

As shown in Fig.\ref{Ex}, the image depicts the propagation of electromagnetic waves during steps $1400$ to $7400$ when the number of particles $n_{pp} = 10^7$, which can reflect the complete process of electromagnetic waves passing through the plasma region. As can be seen from the figure, during the propagation of electromagnetic waves from right to left, there is a clear interaction between the electromagnetic waves and the plasma. By comparing the electromagnetic waves at different times when entering and exiting the plasma region, the attenuation effect of the plasma on the electromagnetic waves can be observed.
Although the obvious attenuation of the electromagnetic wave after entering the plasma can be observed from the Fig.\ref{Ex} , specific values are still needed to be calculated carefully.
\begin{figure}[htbp]
	\centering
	\includegraphics[width=0.5\linewidth]{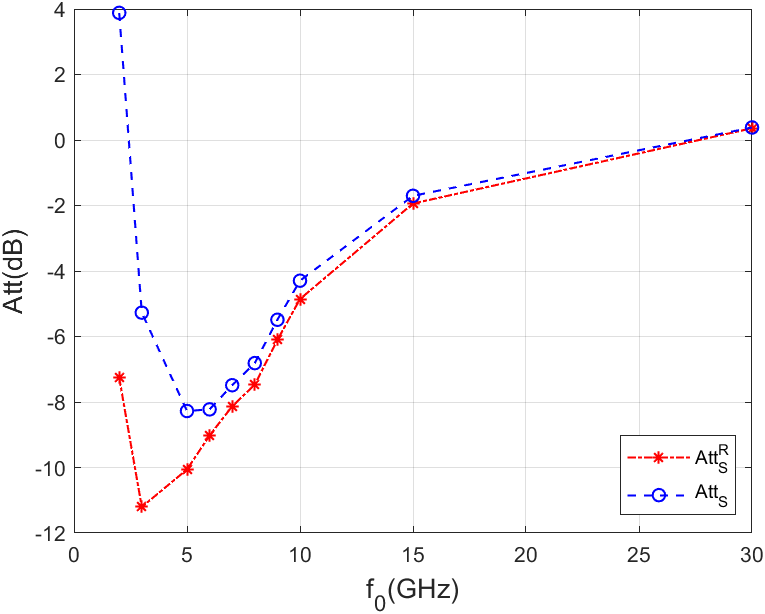}
	\caption{wave attenuation over EM wave frequency.}
        \label{wave_f}
        
\end{figure}

To diagnose the changes in the electric field, we employed ``numerical line probes'' for data collection. Two probes were located at positions $z_i$ and $z_t$, along the $x$ direction. The probe at $z_i$ was situated between $z_2$ and $z_w$, aiming to capture the characteristics of incident and reflected waves; while the probe at $z_t$ was placed between $z_0$ and $z_1$ to monitor the properties of transmitted waves. Through this method, we were able to record in detail the dynamics of the electric field during the simulation. However, the distribution of the circular plasma region led to different degrees of electromagnetic wave attenuation at different $x$ positions. As shown in Fig.\ref{sum:zt}, the Poynting vector $S \equiv |\bm{E} \times \bm{B}| / \mu_0$ diagnosed by the probe at $z_t=800$ was plotted at different times, which is a reasonable quantity as it reflects the power of electromagnetic wave radiation.

The  Fig.\ref{zt_1} shows the transmitted power at each collection during the first half-wave, and the Fig.\ref{zt_2} during the second half-wave. These graphs correspond to the distribution characteristics of the number of particles in the circular plasma region, i.e., higher in the middle and lower on both sides. Additionally, the concave and convex shapes of the two graphs also correspond to the two half-wave cycles of the sinusoidal waveform of the electromagnetic wave. Although the distribution of the circular plasma region led to different degrees of attenuation of electromagnetic waves at different positions of vertical incidence, the average attenuation over the entire computational domain still met the engineering requirements for electromagnetic wave attenuation. Therefore, this model can, to some extent, reflect the attenuation process of the interaction between electromagnetic waves and plasmas in actual scenarios.
\begin{figure}[htb]
	\centering
	\includegraphics[width=0.48\linewidth]{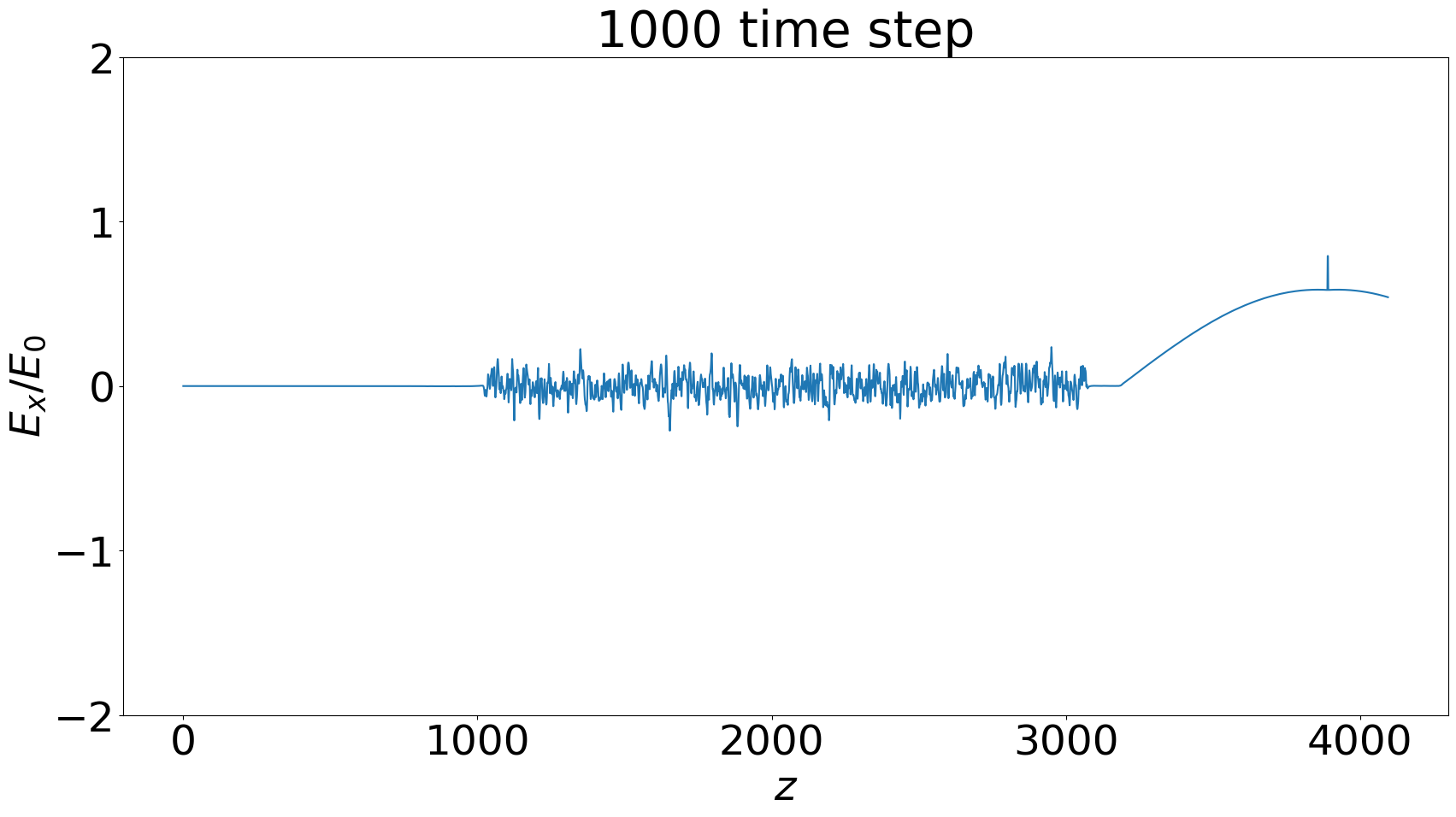}
	\caption{Electromagnetic Wave Propagation Process of the case with $E_0$=$2$×$10^5$.}
	\label{2e5}
\end{figure}

In order to more comprehensively understand the variation in attenuation of electromagnetic waves at different positions due to the distribution of the circular plasma region when they are incident perpendicularly, we plotted the power of electromagnetic wave radiation for grid numbers $x=1536$, $x=1024$, and $x=2048$ in the radial direction, as well as the average value of electromagnetic wave radiation power collected by all probes in Fig.\ref{x=}. Fig.\ref{x=zt} shows the incident electromagnetic wave radiation power at $z_i=3276.8$ , and it can be seen that these four curves almost completely overlap; Fig.\ref{x=zi} shows the transmitted electromagnetic wave radiation power at $z_t=800$, and it can be seen that the trends of these four curves are basically consistent, differing only at three peaks. The curves at $x=1536$ and $x=1024$ are very messy at the peaks, while the curves at $x=2048$ and the average value are very smooth. However, the data values at $x=2048$ in the figure are all greater than the other three curves, which can only represent the place with the fewest particles. Therefore, we choose the average value of the electromagnetic wave radiation power on the probe line for further analysis. $S(z_i)$ clearly shows the two peaks of the incident sinusoidal wave ($S_i$) and the two main peaks of the reflected wave ($S_r$). $S(z_t)$ indicates the transmitted wave ($S_t$). Therefore, we can use the peaks (maximum values) to evaluate the wave attenuation (unit:dB),
\begin{equation}\label{eq:AttS}
    \textrm{Att}_\textrm{S} \equiv 10 \log_{10} \left[
    \dfrac{\textrm{max}(S_t)}{\textrm{max}(S_i)}
    \right].
\end{equation}
We can read
$\textrm{max}(S_i) \approx 2.27 \times 10^6$ w/m$^2$
and
$\textrm{max}(S_t) \approx 1.47 \times 10^6$ w/m$^2$
in Fig.\ref{x=},
thus $\textrm{Att}_\textrm{S} \approx -1.89$ dB.
We can also subtract the reflected waves
and define
\begin{equation}\label{eq:AttSR}
    \textrm{Att}_\textrm{S}^\textrm{R}
    \equiv 10 \log_{10} \left[
    \dfrac{\textrm{max}(S_t)}
    {\textrm{max}(S_i)-\textrm{max}(S_r)}
    \right],
\end{equation}
where we can read
$\textrm{max}(S_r) \approx 1.24 \times 10^5$ w/m$^2$
in Fig.\ref{x=},
thus $\textrm{Att}_\textrm{S}^\textrm{R} \approx -1.64$ dB.
In the subsequent sections, when the two main peaks of the reflected wave are significantly different, we will use the average of the two peaks of the reflected wave as $S_r$ to evaluate the attenuation, rather than the maximum peak.

\section{Effects of physical parameters on wave attenuation}
\label{sec:3}

In this section, we investigate the influence of numerical and  physical parameters on wave attenuation by varying the total number of macro-particles, electromagnetic wave frequency, amplitude, and plasma distribution structure, while keeping other parameters the same as those applied in  Sec.\ref{sec:2}.

\subsection{The impact of macro-particles number}
Firstly, we examine the impact of the number of macro-particles on the simulation. We investigate the effect of the total number of macro-particles $n_{pp}$ on the attenuation of electromagnetic waves in the simulation by varying $n_{pp}$ to $10^6$, $10^7$, and $10^8$, while keeping all other conditions constant. As shown in the left panel of Fig.\ref{sum:npp}, the incident power does not change with the variation in the 
\begin{figure}[htb]
	\centering
	\includegraphics[width=0.5\linewidth]{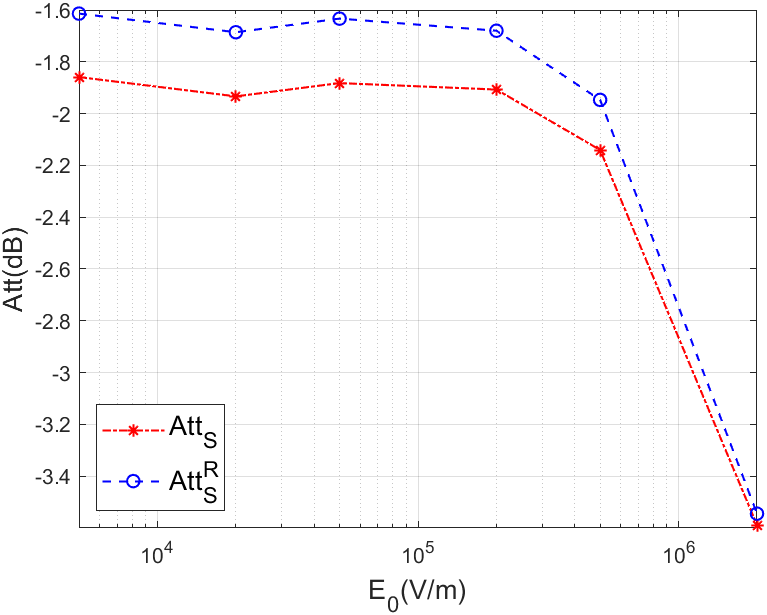}
	\caption{Wave attenuation over EM wave amplitdue.}
	\label{E0}
\end{figure}
number of macro-particles, and the lines for $n_{pp}=10^7$ and $n_{pp}=10^8$ overlap when reflecting, with only a slight difference observed for the data with $n_{pp}=10^6$. The transmitted power through the plasma for the three different particle numbers is shown in the right panel of Fig.\ref{sum:npp}.The lines for $n_{pp}=10^7$ and $n_{pp}=10^8$ are still highly overlapping, while a difference at the peak is observed for $n_{pp}=10^6$ compared to the other two particle numbers.

\begin{figure}[htbp]
	\centering
		\includegraphics[width=0.48\linewidth]{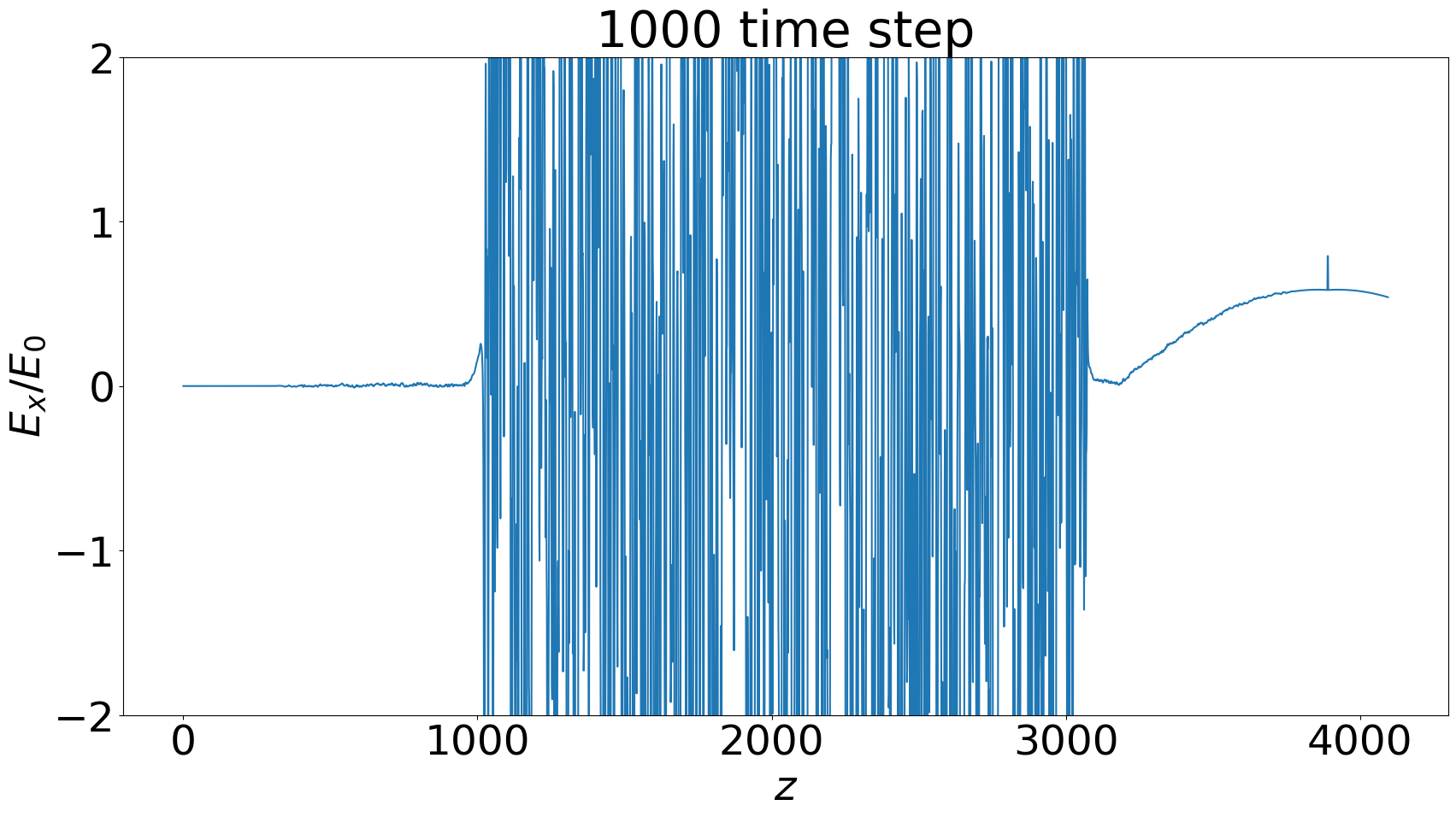}
		\includegraphics[width=0.48\linewidth]{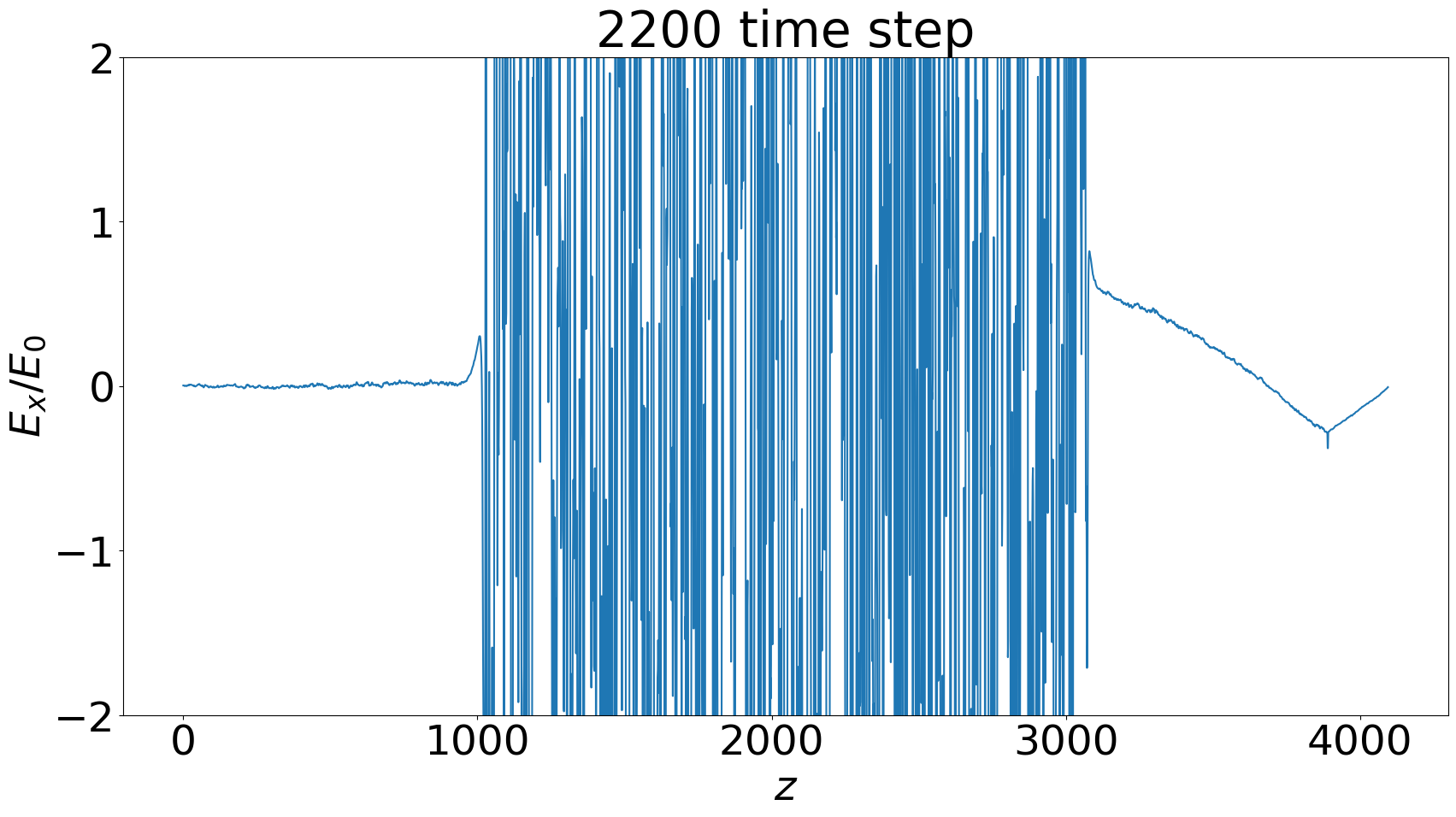}
	\caption{1D Electromagnetic Wave Propagation Process of the case with $E_0$=$5$×$10^3$V/m.}
	\label{5e3}
\end{figure}

When using the Particle-in-Cell (PIC) method to study the interaction between plasmas and electromagnetic waves, it is commonly believed that when a small number of macro-particles are used to address the issue of wave attenuation, the noise displayed in 1D images is too high, making it difficult to discern the attenuation and reflection processes of the electromagnetic waves. However, when a large number of macro-particles are set, although the plasma radiation noise becomes very small, the computational load is significant. As shown in the Fig.\ref{fig:npp1D}, it is clear that the plasma radiation noise is minimal when the number of particles is $n_{pp}=10^8$ and maximum when it is $n_{pp}=10^6$. In actual calculations, a simulation with $n_{pp}=10^6$ particles takes $4$ hours to complete, with $n_{pp}=10^7$ particles it takes $7$ hours, and with $n_{pp}=10^8$ particles, it takes $18$ hours. Therefore, considering the impact of particle number on attenuation and noise, as well as the computational load, it was ultimately determined that a particle number of $n_{pp}=10^7$ would be used for subsequent simulations.

\subsection{Varying EM wave frequency}
This section investigates the impact of EM wave frequency ( $f_0 $) on wave attenuation. Numerous research findings indicate that plasmas exhibit the properties of a high-pass filter. When the frequency of the incident electromagnetic wave is greater than the plasma frequency, the electromagnetic wave can propagate within the plasma, accompanied by energy attenuation, which is consistent with fundamental plasma theory. As the frequency of the electromagnetic wave decreases, the plasma will reflect the wave. When the frequency of the incident electromagnetic wave is less than the plasma frequency, the plasma will produce strong reflection of the wave, and at this point, the reflected power must be subtracted to calculate the attenuation. 
\begin{figure}[htb]
	\centering
	\subfloat[staggered two-row distribution]{
		\includegraphics[width=0.48\linewidth]{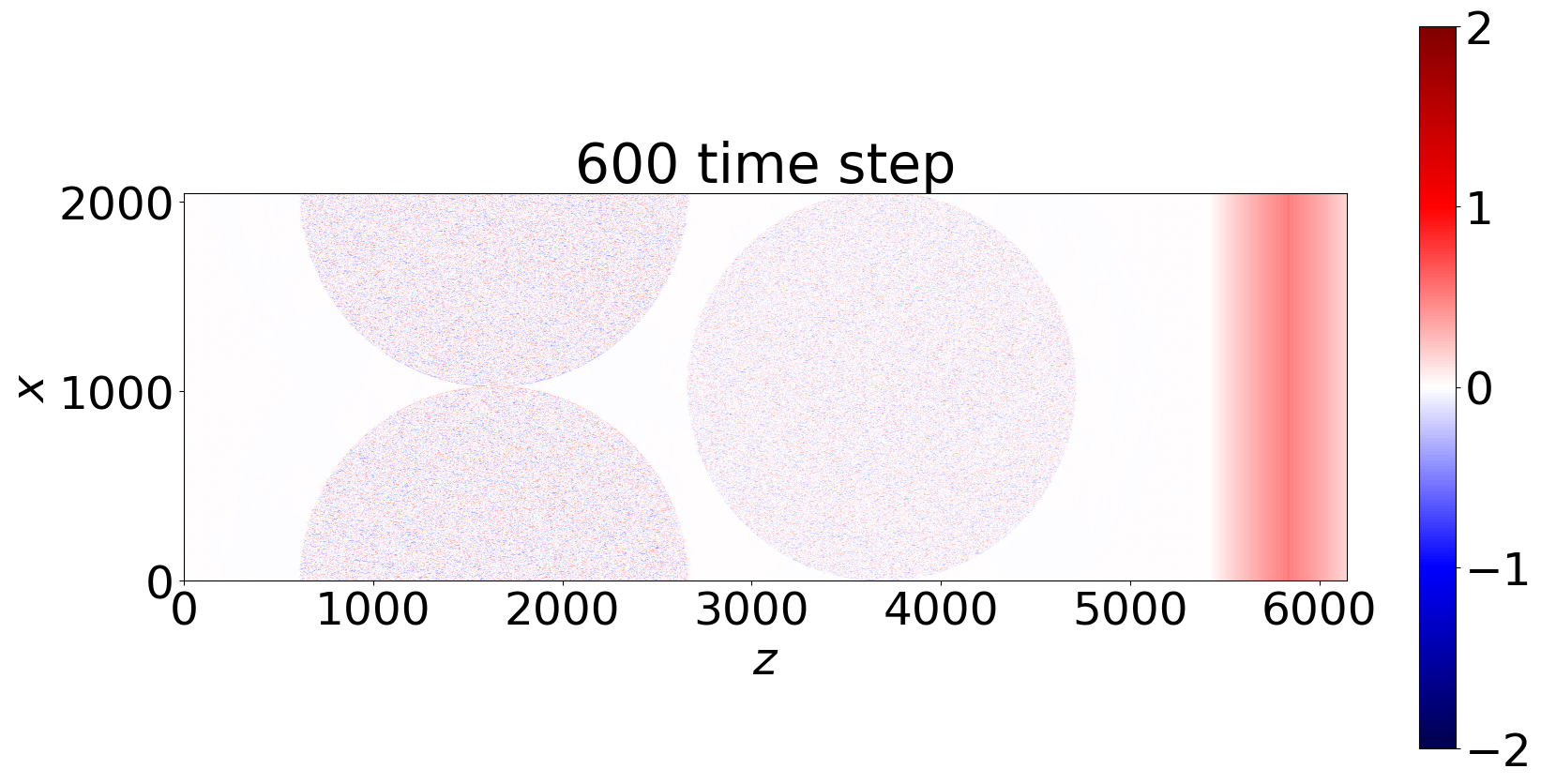}
		\label{s}
	} 
	\subfloat[two-row parallel plasma distribution]{
		\includegraphics[width=0.48\linewidth]{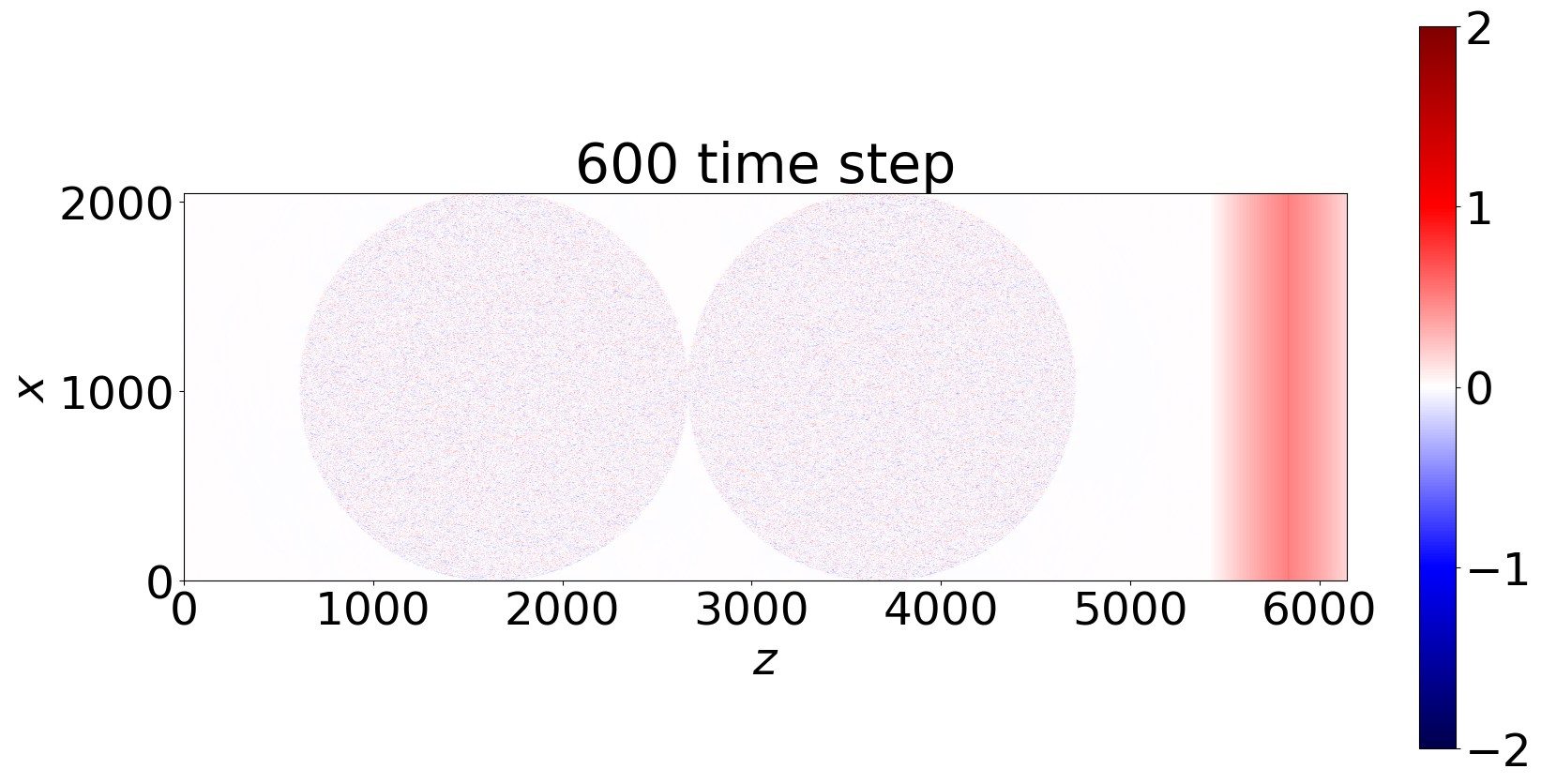}
		\label{p}
	} \\
	\subfloat[single-row plasma distribution]{
		\includegraphics[width=0.48\linewidth]{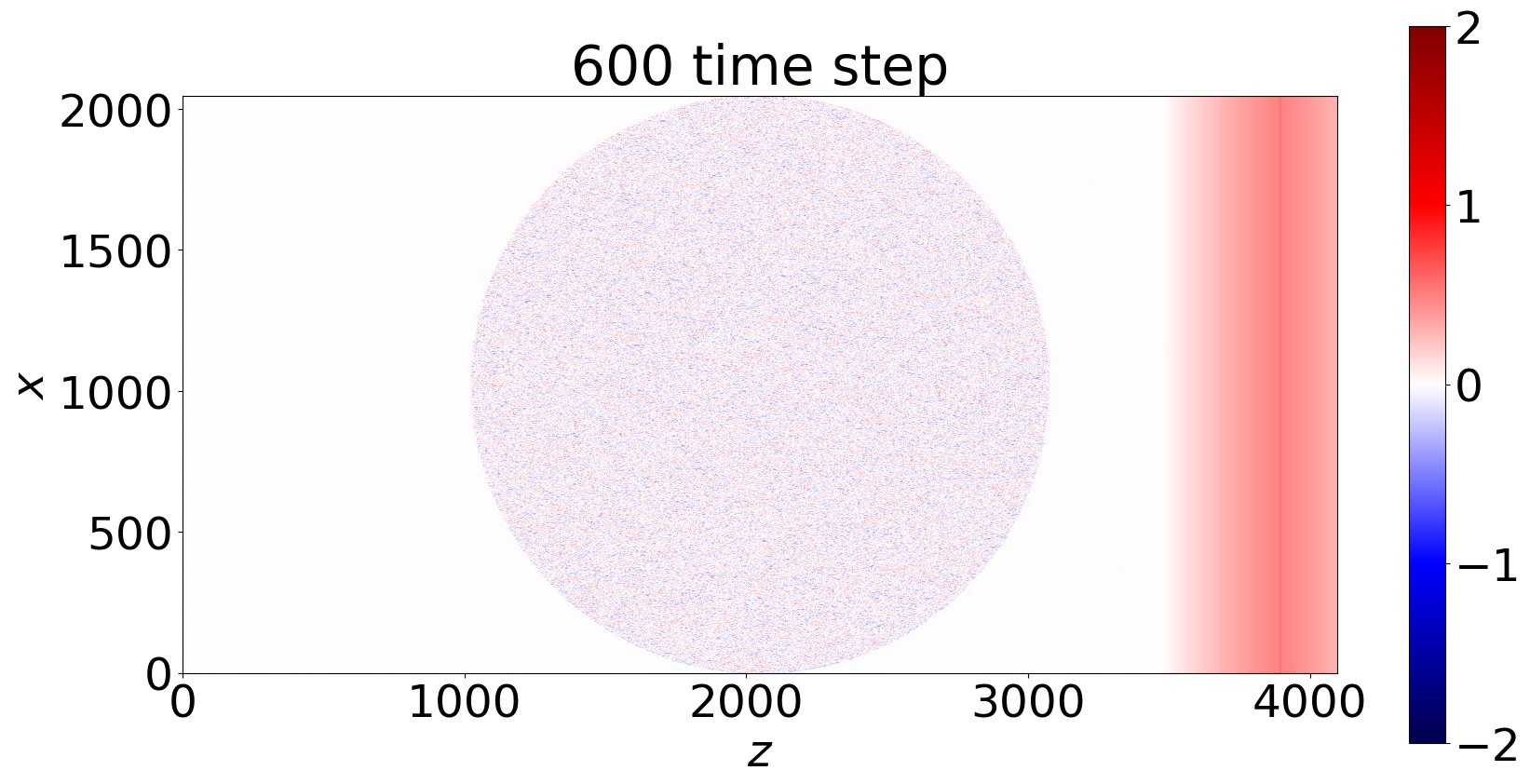}
		\label{sin}
	}
	\caption{Different Plasma Distributions.}
	\label{stru}
\end{figure}

As shown in the Fig.\ref{wave_f}, for a plasma density of $10^{18}$m$^{-3}$and a plasma frequency of $9$ GHz, the relationship between electromagnetic wave attenuation and the frequency of the wave is depicted for different incident electromagnetic wave frequencies. We can observe that when the frequency of the incident electromagnetic wave is low, the attenuation is minimal, particularly the $\textrm{Att}_\textrm{S}^\textrm{R}$line after subtracting the reflected wave is more pronounced. This is because when the electromagnetic wave frequency is less than the plasma frequency, causing total reflection of the wave at the plasma interface, preventing it from entering the plasma interior. As the electromagnetic wave frequency approaches $9$ GHz, reflection decreases, and the amount of electromagnetic wave attenuation significantly increases.
\begin{figure}[htbp]
	\centering
	\includegraphics[width=0.44\linewidth]{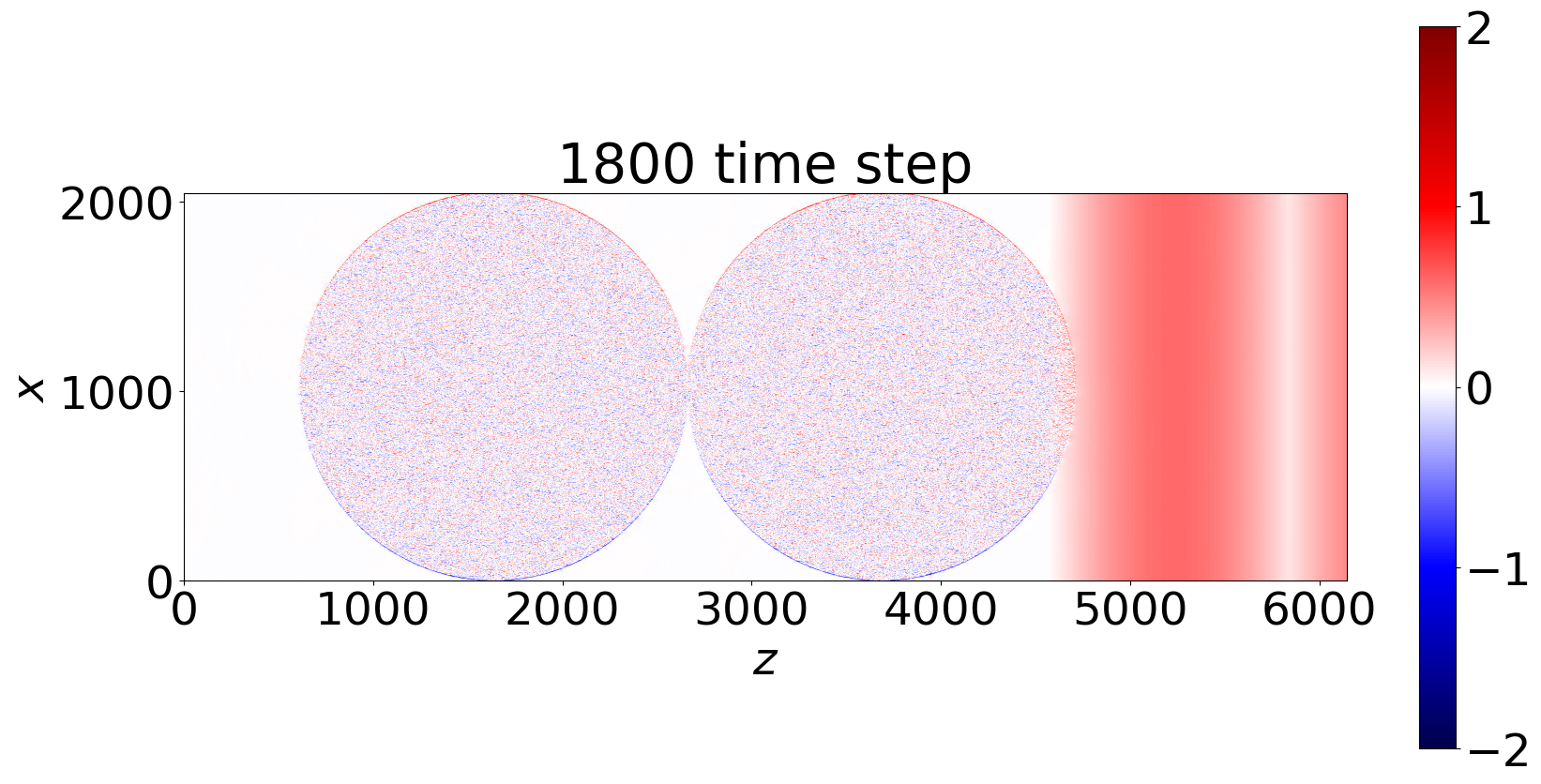}
	\includegraphics[width=0.44\linewidth]{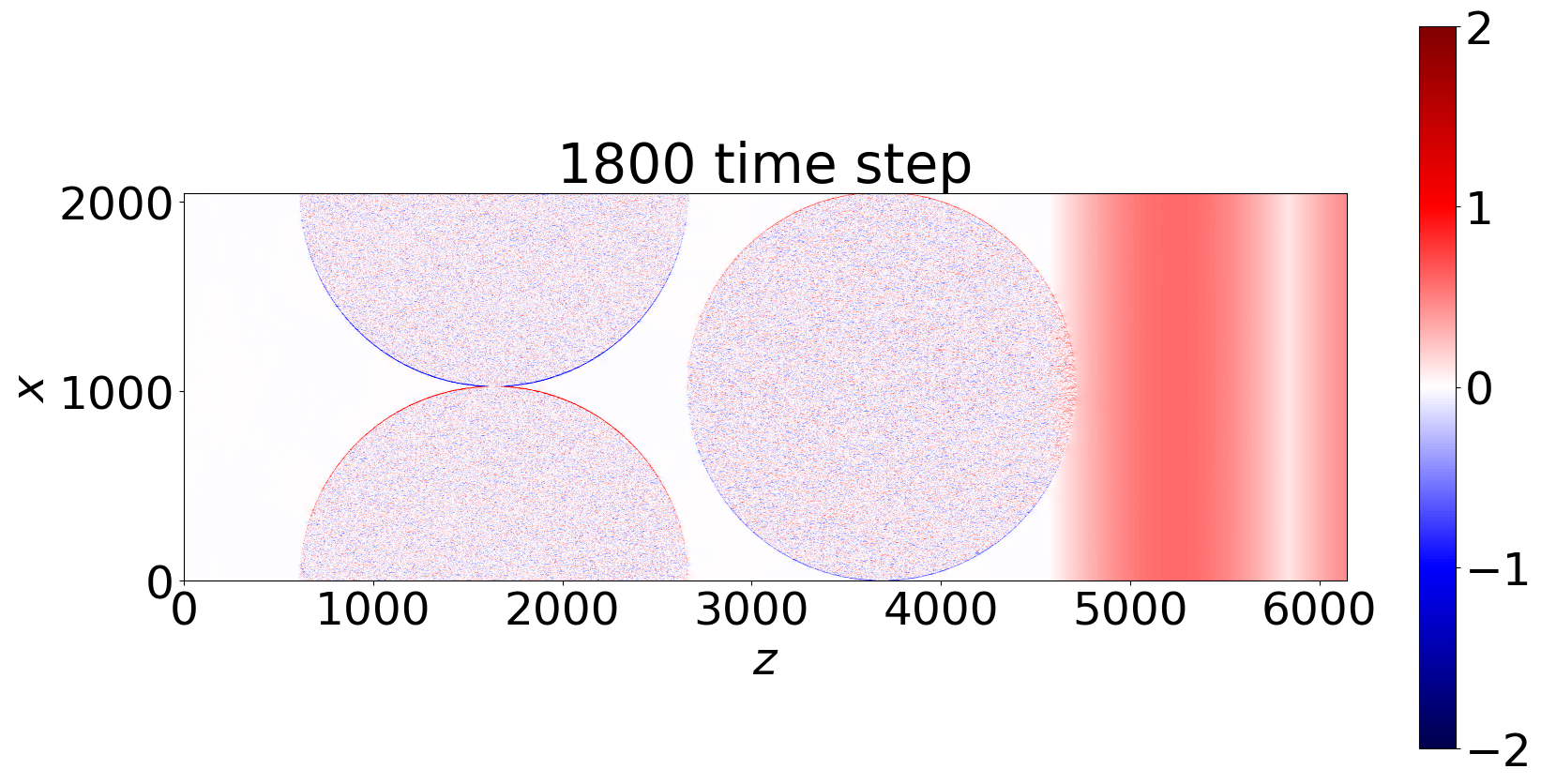}\\
        \includegraphics[width=0.44\linewidth]{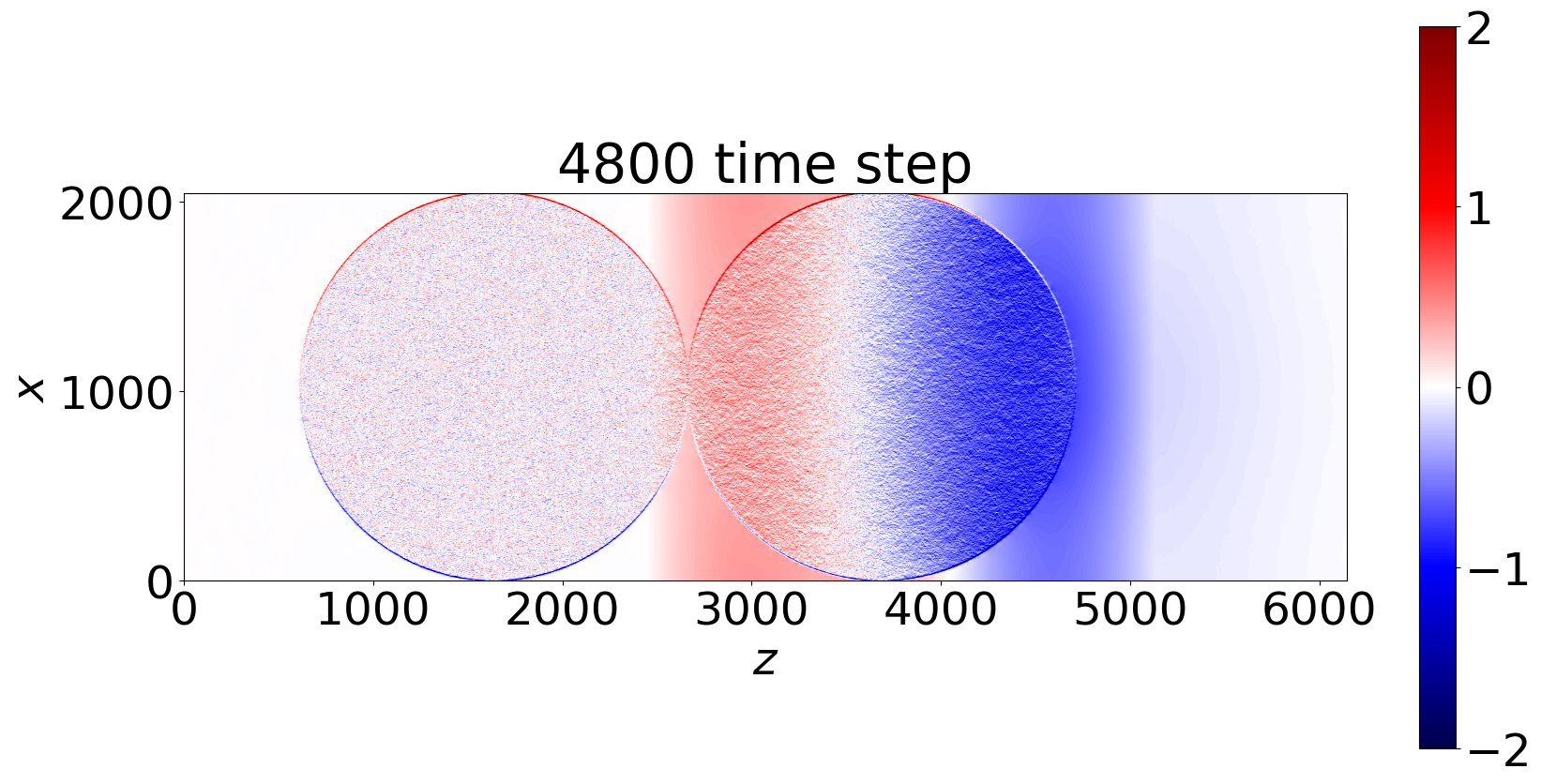} 
	\includegraphics[width=0.44\linewidth]{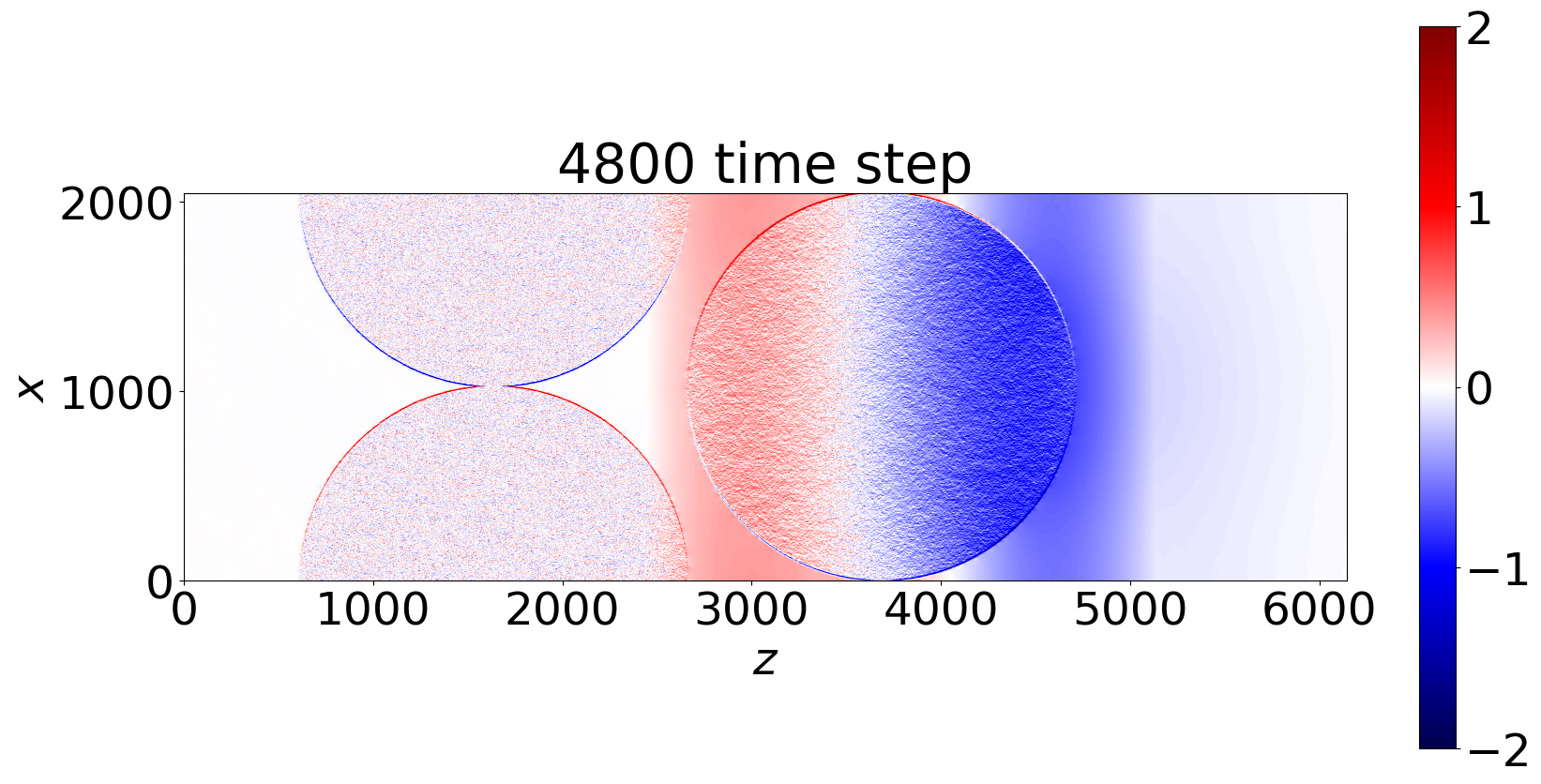}\\
	\subfloat[$2$D Plasma Transport Process of Parallel Plasma Distribution]{
		\includegraphics[width=0.44\linewidth]{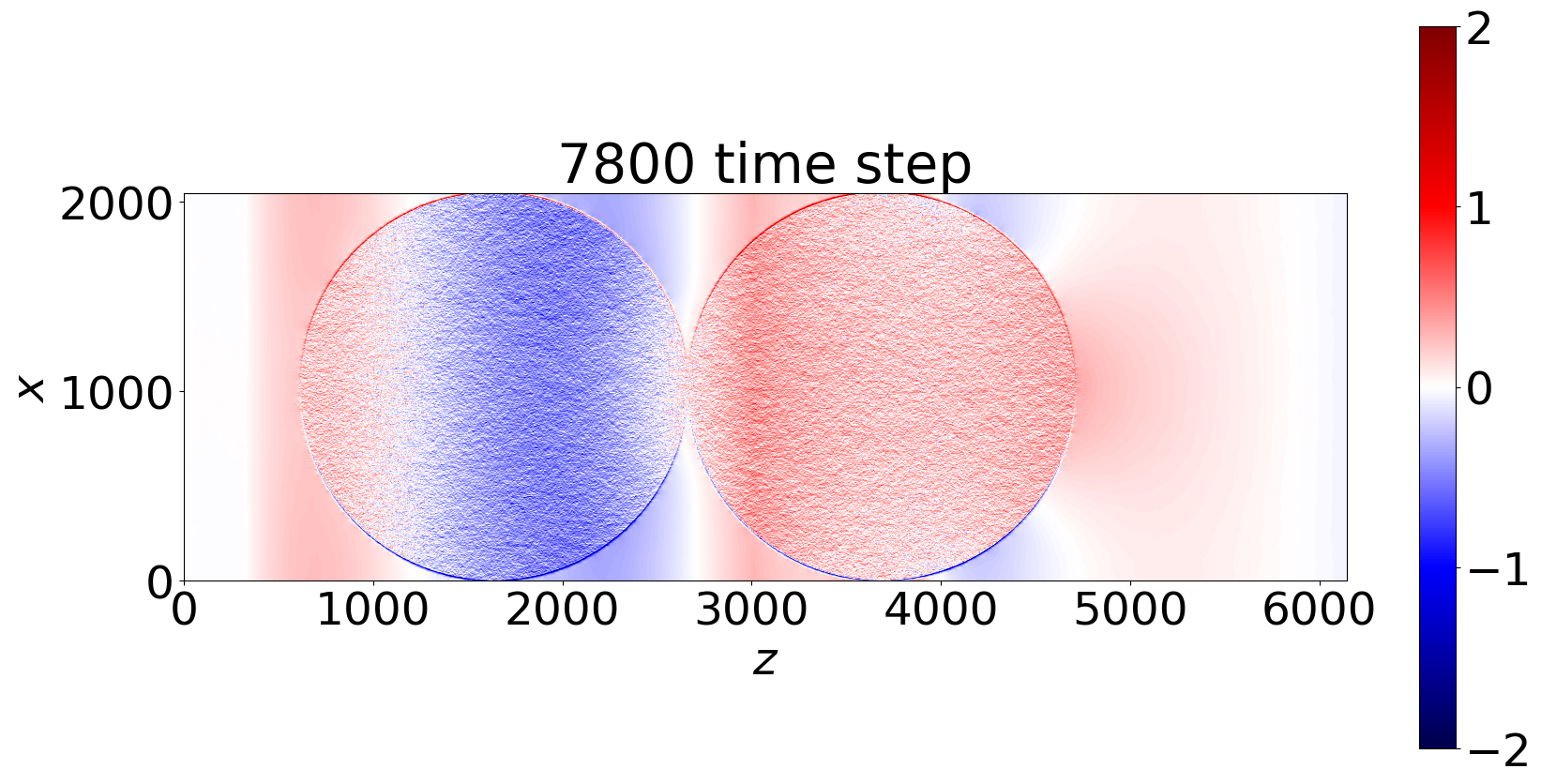}
		\label{pa}
	} 
	\subfloat[$2$D Plasma Transport Process of Parallel Plasma Distribution]{
		\includegraphics[width=0.44\linewidth]{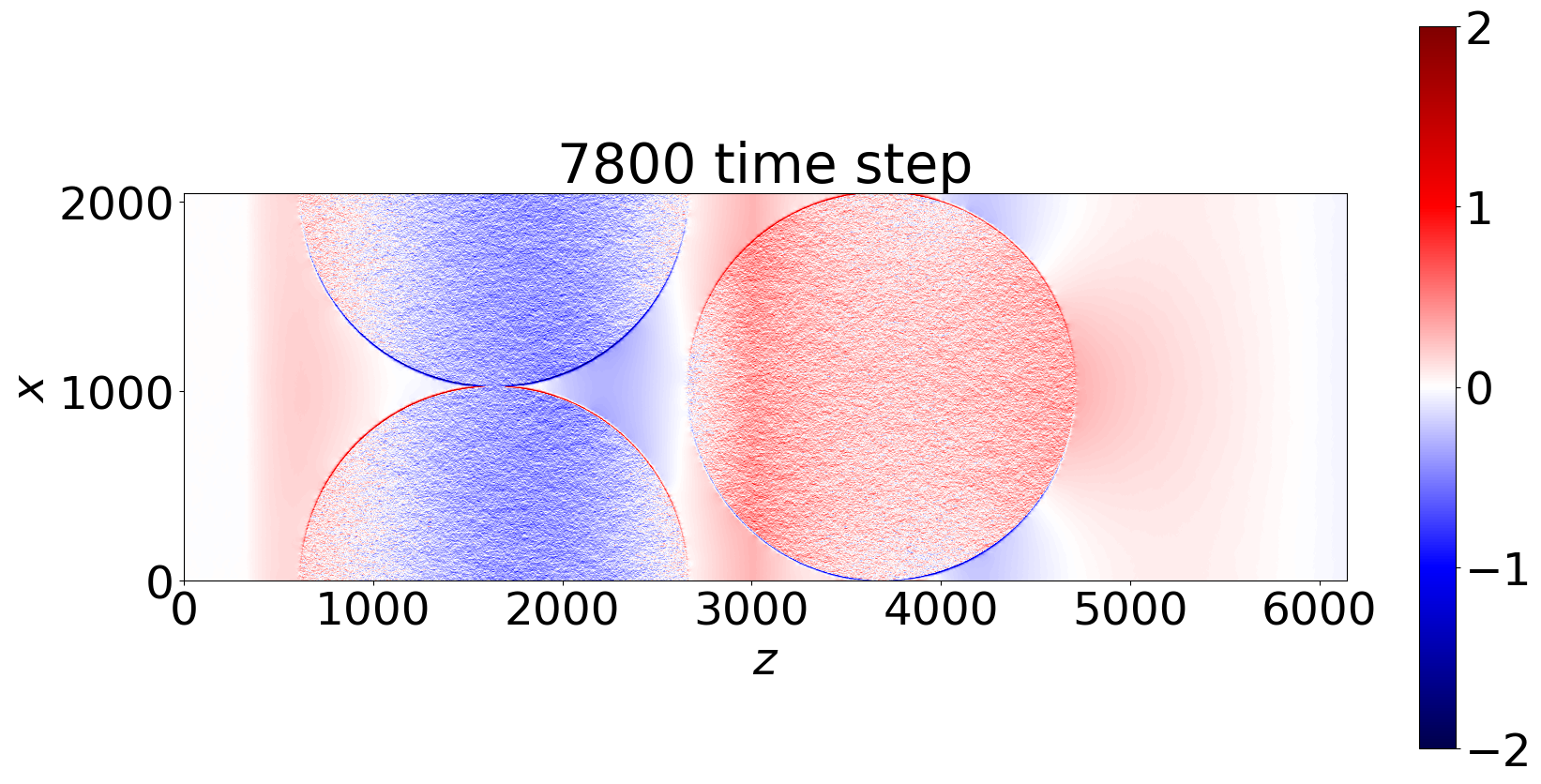}
		\label{st}
	}
	\caption{$2$D Plasma Transport Process with Different Plasma Distributions.}
	\label{pa and st}
\end{figure}
This is due to the plasma frequency being close to the frequency of the incident wave at this point, causing the wave to propagate near the cutoff region and leading to resonant absorption, which can cause substantial attenuation of the electromagnetic wave.
Continuing to increase the frequency of the electromagnetic wave moves it further from the cutoff frequency, and it can be seen that the $\textrm{Att}_\textrm{S}^\textrm{R}$ and $\textrm{Att}_\textrm{S}$ lines gradually converge, with reflection gradually diminishing. The attenuation of the electromagnetic wave by the plasma will then primarily be due to collisional absorption. When the electromagnetic wave frequency is too high, the electrons in the plasma cannot respond quickly enough to the changes in the wave, and they cannot polarize effectively. As a result, their ability to absorb the energy of the electromagnetic wave is greatly reduced, leading to decreased attenuation of the wave within the plasma, making it easier for the wave to penetrate through the plasma.

\subsection{Varying EM wave amplitude }

In the subsequent study, we explored the influence of amplitude on Particle-in-Cell (PIC) simulations. We adopted the amplitude range from $5$×$10^3$V/m to $2$×$10^6$V/m  as referenced in Yan’s research\cite{yan2024numerical} to verify whether the effects of amplitude on the interaction between electromagnetic waves and plasma are minor in two-dimensional simulations as they are in one-dimensional cases. It was found that in 1D simulations, the loaded plasma generates a noisy field initially, and a high-magnitude portion of this noisy field propagates out of the plasma region. However, in 2D simulations studied in this work, the initial noisy field exists but does not propagate much out of the plasma.

Using $E_0$=$2$×$10^5$V/m as a case study, in the Fig.\ref{2e5}, the y-axis represents $E_x$/$2$×$10^5$, where $E_x$ is the actual amplitude of the excitation source. The electromagnetic wave is initially excited at the point $z_w$, which is a different setup from Yan’s where the Electromagnetic noise are transmitted before the wave excitation. This approach can significantly reduce the computational effort and time required for this simulation. Moreover, when assessing decay, it is unnecessary to subtract the energy associated with noise. The Fig.\ref{E0} illustrates the temporal evolution of the Poynting vector. The decay trend aligns with Yan’s findings. For a certain range of $E_0$, the decay remains stable; however, as $E_0$ increases further, the decay rate accelerates due to energy loss from excitation or ionization processes. As shown in the Fig.\ref{5e3}, this is the propagation image of the electromagnetic wave when the amplitude is $5$×$10^3$V/m. When $E_0$ is too small, the electromagnetic wave will be drowned in the plasma noise, affecting the diagnosis.

In summary, due to the interactions between particles within the plasma region, the plasma will continuously vibrate at its location, generating a small Electromagnetic noise, which is the noise signal of the plasma. Due to the presence of this noise signal, it is necessary to select an appropriate amplitude value when setting simulation parameters to ensure that the electromagnetic wave signal is not obscured by the noise signal.

\subsection{Plasma distribution structures }

This section discusses the impact of different plasma distribution structures on electromagnetic wave attenuation. As shown in the Fig.\ref{stru}, the simulation images depict various distribution states with an amplitude of $2$×$10^5$V/m, 
an electromagnetic wave frequency of $15$ GHz, and a certain radius. Fig.\ref{s} shows a staggered two-row distribution where the line connecting the centers of adjacent three discharge tubes can form an equilateral triangle, Fig.\ref{p} shows a two-row parallel plasma distribution, and Fig.\ref{sin}  shows a single-row plasma distribution.
Additionally, in order to reduce the computational load and simulate the absorption effect of a large number of discharge tubes on electromagnetic waves, periodic boundary conditions are applied to both the upper and lower boundaries of the calculation area. This ensures that the plasma region continues to exist outside the calculation area, thereby eliminating the phenomenon of electromagnetic waves bypassing at the farthest boundary due to an excessively large wavelength.

Fig.\ref{pa and st} shows the 2D electromagnetic wave propagation images for parallel and staggered distributions from step 1800 to 7800. This figure can reflect the complete propagation process of electromagnetic waves passing through the plasma region from right to left. It can be seen that the plasma has an attenuating effect on the electromagnetic waves when passing through the plasma region.

\begin{figure}[htb]
	\centering
	\subfloat[Time evolution of the Poynting vectors at $z_i$]{
		\includegraphics[width=0.48\linewidth]{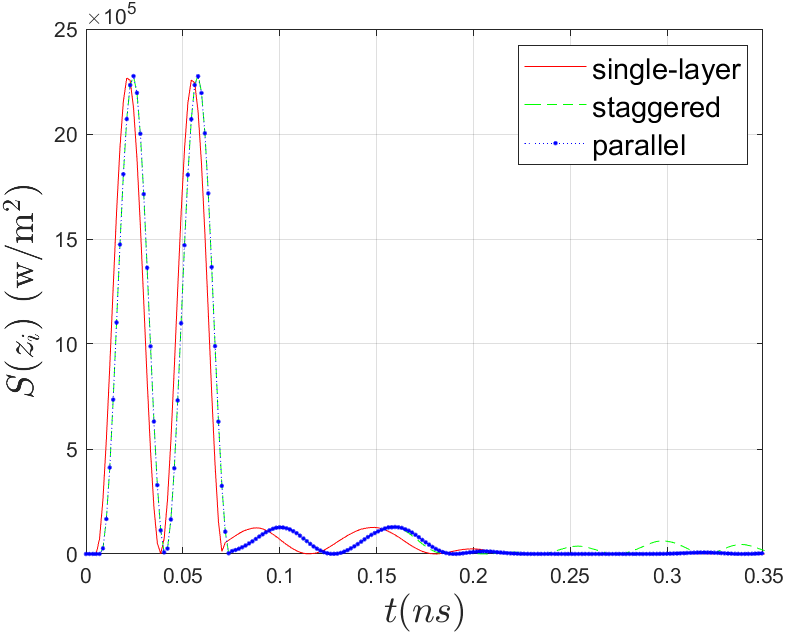}
		\label{zi_stru}
	} \hfill
	\subfloat[Time evolution of the Poynting vectors at $z_t$]{
		\includegraphics[width=0.48\linewidth]{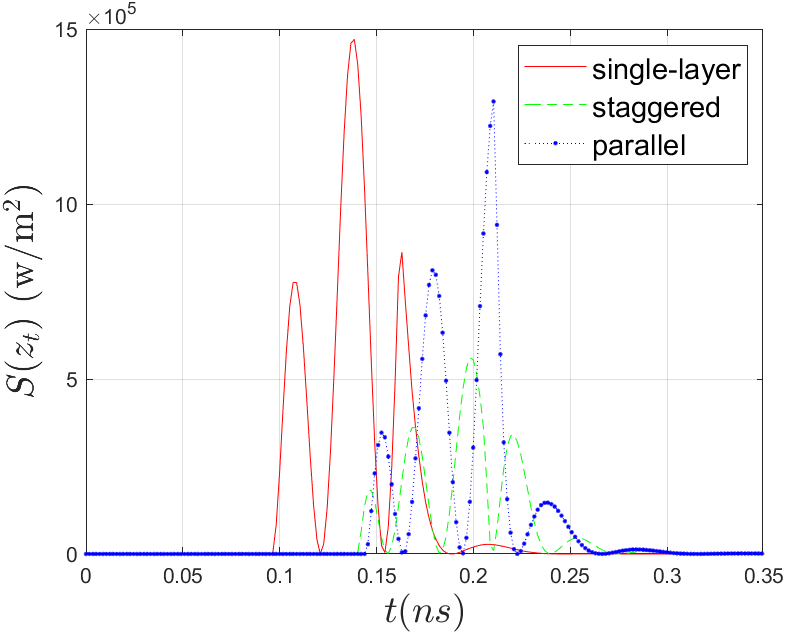}
		\label{zt_stru}
	}
	\caption{Time evolution of the poynting vectors with different plasma distributions.}
	\label{att_stru}
\end{figure}

Comparing the attenuation of electromagnetic waves in three distribution methods, as shown in the Fig.\ref{zi_stru}, it is clear that since the relevant values of the incident electromagnetic waves at the initial time are the same, the power images before entering the plasma for the three distribution states are identical. As shown in the Fig.\ref{zt_stru}, this is the transmitted power of the electromagnetic waves after passing through the plasma region. It can be clearly seen that the attenuation effect of the plasma structure with two staggered rows is greater than that of the plasma distribution with two parallel rows, which in turn is greater than the single-row plasma distribution. This can also be verified by calculations according to the Eq.\eqref{eq:AttSR}.

\subsection{Varying plasma collision types }
\begin{figure}[htb]
	\centering
	\subfloat[Time evolution of the Poynting vectors at $z_i$]{
		\includegraphics[width=0.48\linewidth]{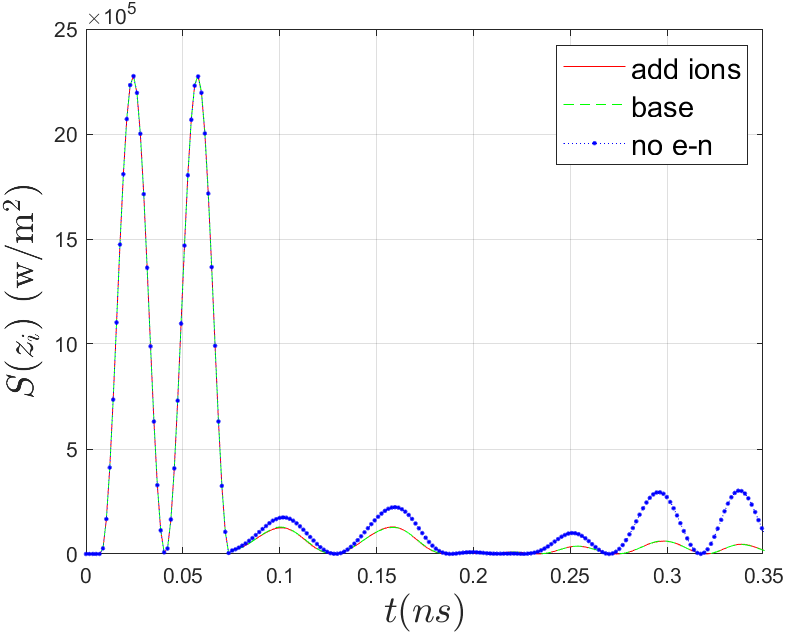}
		\label{eni}
	} \hfill
	\subfloat[Time evolution of the Poynting vectors at $z_i$]{
		\includegraphics[width=0.48\linewidth]{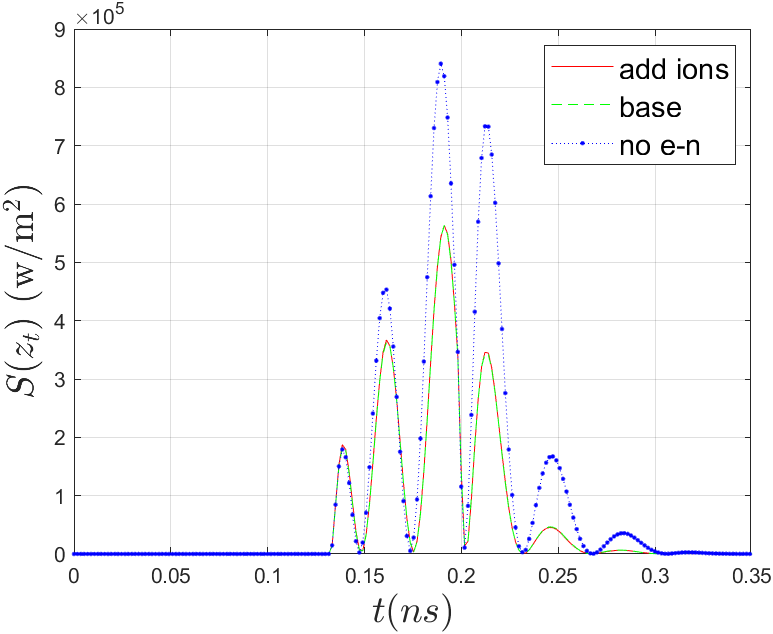}
		\label{ent}
	}
	\caption{Time evolution of the poynting vectors with different  plasma collision types.}
	\label{sum:en}
\end{figure}

The work in this paper utilizes the MCC algorithm to account for collision effects. Previous studies only considered the elastic collisions of electrons, ionization, and excitation, while some collisions were not taken into account. Therefore, in this section, we further investigate the collision effects by using the MCC algorithm to consider different types of collisions. As shown in the Fig.\ref{sum:en}, $1$) simultaneous consideration of elastic collisions, ionization, and excitation of both ions and electrons, labeled as ``add ions''; $2$) no consideration of ion collisions, only electron collisions, labeled as ``base''; $3$) only considering ionization and excitation in electron collisions, labeled as ``no e-n''. As shown in Fig.\ref{eni}, without adding electron elastic scattering, more energy of the electromagnetic wave is reflected by the plasma, and the impact of the collision cases with increased ions on the reflection of electromagnetic waves is small. From Fig.\ref{ent}, it can be seen that the overlap between ``add ions'' and ``base'' is very high, indicating that the collision of ions has little impact on the attenuation of the entire simulation. However, when the electron elastic scattering collision is removed, the amplitude of the curve in the transmission wave diagram is almost twice that of ``add ions'' and ``base''. Therefore, it can be concluded that electron elastic scattering is indeed the most important collision type for electrons, while the impact of collisions related to ions can be ignored because, compared to electrons, ions are almost stationary.

\section{Conclusion}
\label{sec:4}

This paper employs two-dimensional electromagnetic Particle-in-Cell (PIC) simulations to meticulously investigate the impact of numerical parameters in PIC simulations and the influence of physical parameters on wave-plasma interactions. The paper uses the Poynting vector (i.e., power) wave attenuation calculation method to explore the effects of particle number, electromagnetic wave frequency, amplitude, 
collision types, and plasma array distribution on the simulations and electromagnetic wave attenuation. The study finds that the maximum wave attenuation can be achieved when the EM wave frequency is close to the plasma oscillation frequency. For the amplitude of EM waves, within a certain range, the amplitude of the electromagnetic wave is independent of wave attenuation. Appropriately increasing the amplitude of the electromagnetic wave can ensure that the electromagnetic wave signal is not submerged in noise signals and can help quickly locate the plasma region. Selecting an appropriate number of particles and amplitude can make the two-dimensional PIC simulations more efficient and accurate. The results indicate that under the same conditions, the staggered distribution of the plasma array has the best absorption effect,
which agrees with our recent experimental
observation that will be published
in a separate paper.
In subsequent research, we will continue to explore the plasma array and study the impact of non-uniform plasma distributions on the absorption efficiency of electromagnetic waves. We will also compare the simulation results with our experimental measurements.


\section*{Acknowledgment}

This research used the open-source particle-in-cell code WarpX
\url{https://github.com/ECP-WarpX/WarpX},
primarily funded by the US DOE Exascale Computing Project.
Primary WarpX contributors are with LBNL, LLNL, CEA-LIDYL,
SLAC, DESY, CERN, and TAE Technologies.
We acknowledge the contributions of all WarpX contributors and the support from the National Natural Science Foundation of China (Grant Nos. 5247120164 and 52177135).
\bibliographystyle{unsrturl}
\bibliography{reference}

\begin{thebibliography}{10}

\bibitem{SJES6AFFAE5AEEC7647980C0AA1D7ED0010B}
Ouyang Wenchong, Ding Chengbiao, Liu Qi, Lu~Quanming, and Wu~Zhengwei.
\newblock Arrayed multiple atmospheric-pressure plasma jet sources for active
  stealth.
\newblock {\em Cell Reports Physical Science}, 4(12):101715--, 2023.

\bibitem{XJAZB4C513E52A39B80C506EC24707BC1448}
Zhang Qingchao, Tian Zengyao, Tang Wenyuan, Tang Nian, Zhao Hu, and Lin Hui.
\newblock Study of attenuation characteristics of electromagnetic waves in
  multilayer plasma slabs.
\newblock {\em Journal of Applied Physics}, 125(9):094902--094902, 2019.

\bibitem{57528}
R.J. Vidmar.
\newblock On the use of atmospheric pressure plasmas as electromagnetic
  reflectors and absorbers.
\newblock {\em IEEE Transactions on Plasma Science}, 18(4):733--741, 1990.

\bibitem{1970The}
V.~L. Ginzburg.
\newblock The propagation of electromagnetic waves in plasmas.
\newblock {\em International Series of Monographs in Electromagnetic Waves},
  1970.

\bibitem{liu2006wkb}
Shaobin Liu, Tao Zhou, Xiaoxiang He, Yonggang Zhou, and Wei Hong.
\newblock Wkb and fdtd analysis of terahertz band wave propagation in
  unmagnetized plasmas.
\newblock In {\em 2006 7th International Symposium on Antennas, Propagation \&
  EM Theory}, pages 1--4. IEEE, 2006.

\bibitem{shaobin2008wentzel}
Liu Shaobin, Zhou Tao, Liu Meilin, and Hong Wei.
\newblock Wentzel-kramer-brillouin and finite-difference time-domain analysis
  of terahertz band electromagnetic characteristics of target coated with
  unmagnetized plasma.
\newblock {\em Journal of Systems Engineering and Electronics}, 19(1):15--20,
  2008.

\bibitem{2005Three}
B.~Chaudhury and S.~Chaturvedi.
\newblock Three-dimensional computation of reduction in radar cross section
  using plasma shielding.
\newblock {\em IEEE Transactions on Plasma Science}, 33(6):2027--2034, 2005.

\bibitem{2006Comparison}
Bhaskar Chaudhury and Shashank Chaturvedi.
\newblock Comparison of wave propagation studies in plasmas using
  three-dimensional finite-difference time-domain and ray-tracing methods.
\newblock {\em Physics of Plasmas}, 13(12):470--475, 2006.

\bibitem{2009Study}
B.~Chaudhury and S.~Chaturvedi.
\newblock Study and optimization of plasma-based radar cross section reduction
  using three-dimensional computations.
\newblock {\em IEEE Transactions on Plasma Science}, 37(11):2116--2127, 2009.

\bibitem{fermi1955studies}
Enrico Fermi, P~Pasta, Stanislaw Ulam, and Mary Tsingou.
\newblock Studies of the nonlinear problems.
\newblock Technical report, Los Alamos National Laboratory (LANL), Los Alamos,
  NM (United States), 1955.

\bibitem{harlow1988pic}
Francis~H Harlow.
\newblock Pic and its progeny.
\newblock {\em Computer Physics Communications}, 48(1):1--10, 1988.

\bibitem{xu2013pic}
Yanxia Xu, Xin Qi, Xue Yang, Chao Li, Xiaoying Zhao, Wenshan Duan, and Lei
  Yang.
\newblock Pic-mcc simulation of electromagnetic wave attenuation in partially
  ionized plasmas.
\newblock {\em Plasma Sources Science and Technology}, 23(1):015002, 2013.

\bibitem{gao2022attenuation}
Dong-Ning Gao, Shu-Mei Lin, and Wen-Shan Duan.
\newblock Attenuation of electromagnetic waves in an unmagnetized collisionless
  plasma by particle-in-cell method.
\newblock {\em The European Physical Journal Special Topics},
  231(22):4143--4147, 2022.

\bibitem{xuesong2022investigation}
DENG Xuesong, DING Chenglong, WANG Yahui, LI~Zhigang, Li~Cheng, CHEN Zongsheng,
  LV~Xiangyin, and SHI Jiaming.
\newblock Investigation into wideband electromagnetic stealth device based on
  plasma array and radar-absorbing materials.
\newblock {\em Plasma Science and Technology}, 24(11):114006, 2022.

\bibitem{anderson2007plasma}
Ted Anderson, Igor Alexeff, James Raynolds, Esmaeil Farshi, Sriram
  Parameswaran, Eric~P Pradeep, and Jyothi Hulloli.
\newblock Plasma frequency selective surfaces.
\newblock {\em IEEE Transactions on Plasma Science}, 35(2):407--415, 2007.

\bibitem{danilov1997electromagnetic}
AV~Danilov, SA~Ilchenko, AT~Kunavin, AV~Markov, DV~Sapozhnikov, VY~Yakovlev,
  VA~Permyakov, SN~Tsemko, and VA~Volsky.
\newblock Electromagnetic waves scattering by periodic plasma structure.
\newblock {\em Physica A: Statistical Mechanics and its Applications},
  241(1-2):226--230, 1997.

\bibitem{1997Electromagnetic}
V.~A., Danilov, , , A.~S., Ilchenko, , , T.~A., Kunavin, , , and V.~and A.
\newblock Electromagnetic waves scattering by periodic plasma structure.
\newblock {\em Physica A: Statistical Mechanics and its Applications},
  241(1-2):226--230, 1997.

\bibitem{2005Time}
Mostofa~K. Hpwlader, Yunqiang Yang, and J.~Reece Roth.
\newblock Time-resolved measurements of electron number density and collision
  frequency for a fluorescent lamp plasma using microwave diagnostics.
\newblock {\em IEEE Transactions on Plasma Science}, 33(3):p.1093--1099, 2005.

\bibitem{SIPD42CCED5A681F923759E40183B30D73C4}
B~Wang, J~A Rodríguez, and M~A Cappelli.
\newblock 3d woodpile structure tunable plasma photonic crystal.
\newblock {\em Plasma Sources Science and Technology}, 28(2):02LT01--02LT01,
  2019.

\bibitem{yan2024numerical}
Yize Yan, Fei Du, Jingfeng Tang, Daren Yu, and Yinjian Zhao.
\newblock Numerical study on wave attenuation via 1d fully kinetic
  electromagnetic particle-in-cell simulations.
\newblock {\em Plasma Sources Science and Technology}, 33(11):115013, 2024.

\bibitem{warpx}
Luca Fedeli, Axel Huebl, France Boillod-Cerneux, Thomas Clark, Kevin Gott,
  Conrad Hillairet, Stephan Jaure, Adrien Leblanc, Rémi Lehe, Andrew Myers,
  Christelle Piechurski, Mitsuhisa Sato, Neïl Zaim, Weiqun Zhang, Jean-Luc
  Vay, and Henri Vincenti.
\newblock Pushing the frontier in the design of laser-based electron
  accelerators with groundbreaking mesh-refined particle-in-cell simulations on
  exascale-class supercomputers.
\newblock In {\em SC22: International Conference for High Performance
  Computing, Networking, Storage and Analysis}, pages 1--12, 2022.
\newblock \href {https://doi.org/10.1109/SC41404.2022.00008}
  {\path{doi:10.1109/SC41404.2022.00008}}.

\bibitem{ZhaoPOP2020}
Y.~Zhao, R.~Lehe, A.~Myers, M.~Thevenet, A.~Huebl, C.~B. Schroeder, and J.-L.
  Vay.
\newblock Modeling of emittance growth due to coulomb collisions in
  plasma-based accelerators.
\newblock {\em Physics of Plasmas}, 27(11):113105, 2020.
\newblock \href {https://doi.org/10.1063/5.0023776}
  {\path{doi:10.1063/5.0023776}}.

\bibitem{ZhaoPOP2022}
Y.~Zhao, R.~Lehe, A.~Myers, M.~Thevenet, A.~Huebl, C.~B. Schroeder, and J.-L.
  Vay.
\newblock Plasma electron contribution to beam emittance growth from coulomb
  collisions in plasma-based accelerators.
\newblock {\em Physics of Plasmas}, 29(10):103109, 2022.
\newblock \href {https://doi.org/10.1063/5.0102919}
  {\path{doi:10.1063/5.0102919}}.

\bibitem{Fedeli_2022}
Luca Fedeli, Neïl Zaïm, Antonin Sainte-Marie, Maxence Thévenet, Axel Huebl,
  Andrew Myers, Jean-Luc Vay, and Henri Vincenti.
\newblock Picsar-qed: a monte carlo module to simulate strong-field quantum
  electrodynamics in particle-in-cell codes for exascale architectures.
\newblock {\em New Journal of Physics}, 24(2):025009, mar 2022.
\newblock \href {https://doi.org/10.1088/1367-2630/ac4ef1}
  {\path{doi:10.1088/1367-2630/ac4ef1}}.

\bibitem{Klion_2023}
Hannah Klion, Revathi Jambunathan, Michael~E. Rowan, Eloise Yang, Donald
  Willcox, Jean-Luc Vay, Remi Lehe, Andrew Myers, Axel Huebl, and Weiqun Zhang.
\newblock Particle-in-cell simulations of relativistic magnetic reconnection
  with advanced maxwell solver algorithms.
\newblock {\em The Astrophysical Journal}, 952(1):8, jul 2023.
\newblock \href {https://doi.org/10.3847/1538-4357/acd75b}
  {\path{doi:10.3847/1538-4357/acd75b}}.

\bibitem{Wang_2023}
Baisheng Wang, Tianhang Meng, Yinjian Zhao, Zhongxi Ning, Hui Liu, and Daren
  Yu.
\newblock Electromagnetic particle-in-cell simulation on self-induced magnetic
  field by hollow cathode discharge.
\newblock {\em Plasma Sources Science and Technology}, 32(9):095009, sep 2023.
\newblock \href {https://doi.org/10.1088/1361-6595/acf7e7}
  {\path{doi:10.1088/1361-6595/acf7e7}}.

\bibitem{xie2024effect}
Lihuan Xie, Xin Luo, Zhijun Zhou, and Yinjian Zhao.
\newblock Effect of plasma initialization on 3d pic simulation of hall thruster
  azimuthal instability.
\newblock {\em Physica Scripta}, 99(9):095602, 2024.

\bibitem{Xu_2014}
Yanxia Xu, Xin Qi, Xue Yang, Chao Li, Xiaoying Zhao, Wenshan Duan, and Lei
  Yang.
\newblock \text{PIC-MCC} simulation of electromagnetic wave attenuation in
  partially ionized plasmas.
\newblock {\em Plasma Sources Science and Technology}, 23(1):015002, dec 2013.
\newblock \href {https://doi.org/10.1088/0963-0252/23/1/015002}
  {\path{doi:10.1088/0963-0252/23/1/015002}}.

\bibitem{PANCHESHNYI2012148}
S.~Pancheshnyi, S.~Biagi, M.C. Bordage, G.J.M. Hagelaar, W.L. Morgan, A.V.
  Phelps, and L.C. Pitchford.
\newblock \text{The LXCat} project: Electron scattering cross sections and
  swarm parameters for low temperature plasma modeling.
\newblock {\em Chemical Physics}, 398:148--153, 2012.
\newblock \href
  {https://doi.org/https://doi.org/10.1016/j.chemphys.2011.04.020}
  {\path{doi:https://doi.org/10.1016/j.chemphys.2011.04.020}}.

\end{thebibliography}

\end{document}